\documentclass[12pt]{article}
\usepackage[utf8]{inputenc}
\usepackage{amsmath}
\usepackage{amssymb}
\usepackage{graphicx}

\newcommand{\del}{\partial}
\makeatletter

\DeclareTextSymbolDefault{\textquotedbl}{T1}

\numberwithin{equation}{section}

\@ifundefined{date}{}{\date{}}
\usepackage{braket}
\usepackage{bm}
\usepackage{bbm}
\usepackage{epsfig}
\usepackage{pstricks}
\usepackage{color}
\usepackage{tikz}
\usepackage{lipsum}  

\usepackage{hyperref}

\newcommand{\mat}[1]{\left(\begin{matrix}#1\end{matrix}\right)}

\newcommand{\bbR}{{\mathbb R}}
\newcommand{\bbC}{{\mathbb C}}
\newcommand{\comment}[1]{}

\@ifundefined{showcaptionsetup}{}{%
 \PassOptionsToPackage{caption=false}{subfig}}
\usepackage{subfig}
\makeatother

\begin{document}
\begin{titlepage}
\renewcommand{\thefootnote}{\fnsymbol{footnote}}

\begin{flushright} 
%\today\\
% WIS/06/09-MAY-DPP\\ 
  KEK-TH-2538
\end{flushright} 

%\vspace{0.1cm}
%\vspace{1cm}
\vspace{0.5cm}

\begin{center}
  {\bf \large A new picture of quantum tunneling
%  {\bf \large New insights into quantum tunneling 
    in the real-time path integral
    from Lefschetz thimble calculations}
%  {\bf \large Thimble simulation of real-time quantum tunneling}
  %% {\bf \large A fast Lefschetz thimble calculation\\
  %%   based on backpropagating Hybrid Monte Carlo}
  %%
  %% {\bf \large A simple Hybrid Monte Carlo algorithm\\
%% for the generalized Lefschetz thimble method}
%%
%% {\bf \large Complex Langevin analysis of 
%% the spontaneous breaking of rotational symmetry 
%% in the Euclidean IKKT matrix model}
\end{center}

\vspace{1cm}

%%%%%

\begin{center}
         Jun N{\sc ishimura}$^{1,2)}$\footnote
          { E-mail address : jnishi@post.kek.jp},
         Katsuta S{\sc akai}$^{1,3)}$\footnote
          { E-mail address : sakai.las@tmd.ac.jp} and
         Atis Y{\sc osprakob}$^{1,4)}$\footnote
          { E-mail address : ayosp@phys.sc.niigata-u.ac.jp}
%%%%%

%\maketitle

%\vspace{1cm}
\vspace{1cm}

$^{1)}$\textit{KEK Theory Center,
Institute of Particle and Nuclear Studies,}\\
{\it High Energy Accelerator Research Organization,\\
  1-1 Oho, Tsukuba, Ibaraki 305-0801, Japan}

%~

%% $^{2)}$\textit{Particle and Nuclear Physics Program,
%% Graduate University for Advanced Studies (SOKENDAI),\\
%% 1-1 Oho, Tsukuba, Ibaraki 305-0801, Japan} 

%$^{2)}$\textit{Particle and Nuclear Physics Program,}\\
%\textit{School of High Energy Accelerator Science,}\\
$^{2)}$\textit{Graduate Institute for Advanced Studies, SOKENDAI,\\
1-1 Oho, Tsukuba, Ibaraki 305-0801, Japan} 

$^{3)}$\textit{College of Liberal Arts and Sciences, 
Tokyo Medical and Dental University,}\\
{\it Ichikawa, Chiba 272-0827, Japan}

%~

$^{4)}$\textit{Department of Physics, Niigata University,}\\
{\it
8050 Igarashi 2-no-cho, Nishi-ku, Niigata-shi, Niigata 950-2181, Japan}
\end{center}

\vspace{0.5cm}
%\vspace{1cm}

\begin{abstract}
  \noindent
It is well known that quantum tunneling can be described 
by instantons in the imaginary-time path integral formalism.
However, its description in the real-time path integral formalism
%has not been completely understood.
has been elusive.
%% For instance, if one makes an analytic continuation
%% of an instanton configuration from the imaginary time to the real time,
%% one obtains a singular complex trajectory that wind 
%% infinitely many times around the 
%% potential minima in the complex plane.
% partly due to the problem of oscillatory integrals. 
Here we 
%use the Picard-Lefschetz theory to make the oscillatory real-time path integral
%well defined and 
establish a statement that quantum tunneling can be characterized
in general by the contribution of complex 
saddle points, which can be identified
% and the associated thimbles, which can be identified
%the complex saddle points
%which are relevant from the viewpoint of
by using the Picard-Lefschetz theory.
% and the associated thimbles.
%solution to the classical equation of motion.
%in the semi-classical limit $\hbar \rightarrow 0$.
We demonstrate this explicitly by performing Monte Carlo simulations
of simple quantum mechanical systems,
overcoming the sign problem by the generalized Lefschetz thimble method.
%In particular,
We confirm numerically that
the contribution of complex saddle points manifests itself
in a complex ``weak value'' of the Hermitian coordinate operator $\hat{x}$
evaluated at time $t$,
which is a physical quantity that can be measured by experiments in principle.
We also discuss the transition to classical dynamics based on our picture.
%%We also discuss the quantum-to-classical transition along this line.
%%
%% By introducing a sufficiently large momentum in the initial wave function,
%% we reproduce the transition to classical dynamics correctly as well.
%%
%% In particular, we identify the complex saddle points
%% by performing 
%% Monte Carlo simulations
%% of the real-time quantum mechanics with a double-well potential,
%% overcoming the sign problem by the generalized Lefschetz thimble method.
\end{abstract}
\vfill
\end{titlepage}
\vfil\eject

%\newpage

\setcounter{footnote}{0}

\section{Introduction}

Quantum tunneling has been conventionally described by
instantons in the imaginary-time path integral
formalism \cite{coleman_1985,ColemanPaperI,ColemanPaperII},
which enables us to 
%calculate the decay rate of a false vacuum 
investigate, for instance, the decay of a false vacuum 
in quantum field theory
and in quantum cosmology 
within the semi-classical approximation \cite{ColemanPaperI,ColemanPaperII,Abel:2020qzm,Markkanen:2018pdo,Stone:1975bd,Frampton:1976kf,Stone:1976qh,Frampton:1976pb}.
The tunneling amplitude one obtains in this way is suppressed in general
by $e^{-S_0/\hbar}$ with $S_0$ being the Euclidean action
for the instanton configuration,
which reveals its genuinely nonperturbative nature.
%and $\hbar$ is the reduced
%Such techniques are also 
%bubble nucreation in first-order phase transitions,
%domain wall fusions,

Despite this success,
%of the imaginary-time path integral formalism,
it should be noted that such calculations
do not tell us 
%in order to discuss 
%the real-time evolution of 
%
%about the quantum state during and after the tunneling process.
how the tunneling actually occurs.
For that purpose,
%some efforts have been made in the literature
it is important to understand quantum tunneling
in the real-time path integral formalism,
in which it is widely recognized that complex solutions
to the classical equation of motion
play a crucial role.\footnote{See, for instance,
  Ref.~\cite{PhysRevE.68.056211} for
a detailed analysis of complex solutions in a quantum chaos system.}
However, a complete understanding has been missing so far.
For instance, infinitely many complex solutions for a finite elapsed time
have been obtained in simple quantum mechanical 
systems \cite{Turok:2013dfa,Tanizaki:2014xba},
but it was not possible to identify the relevant ones
from the viewpoint of the Picard-Lefschetz theory as we explain shortly.
It was also pointed out that 
the complex trajectories that 
can be obtained by analytic continuation of the instanton solution
%describe quantum tunneling
%should have 
has a spiral shape in the complex plane,
which extends very far from the potential
minimum \cite{Turok:2013dfa,Tanizaki:2014xba}
and becomes singular in the strict real-time limit \cite{Cherman:2014sba}.
%%
%% ones that describe
%% the quantum tunneling from first principles.
%% It was also pointed out that, if one makes an analytic continuation
%% of an instanton solution from imaginary to real time
%% in the case of a double-well potential,
%% one obtains a singular complex trajectory that winds 
%% infinitely many times around the 
%% potential minima in the complex plane \cite{Cherman:2014sba}.

The origin of complex trajectories 
can be naturally understood
in the Picard-Lefschetz theory \cite{Tanizaki:2014xba},
% \cite{Picard1897,Lefschetz1924},
which renders
the oscillatory integral that appears in the real-time
path integral formalism absolutely convergent
by deforming the integration contour
into the complex plane
using the anti-holomorphic gradient flow equation.
%\cite{Alexandru:2015sua}.
Based on Cauchy's theorem,
%In this way, 
one can then rewrite the original integral
as a sum over integrals along the steepest descent contours
(``Lefschetz thimbles'')
associated with some saddle points.
Thus 
%the Picard-Lefschetz 
this theory tells us
% in principle
which saddle points are ``relevant'' to the original path integral.
%Note also that this argument does not rely on the semi-classical 
%approximation.
The problem, however, was that it was technically difficult
to identify the relevant saddle points.

More recently, there have been various developments
on the description of quantum tunneling in the
real-time path integral formalism.
For instance, the optical theorem has been used to
demonstrate that the decay rate of a false vacuum can be correctly reproduced
including one-loop corrections by 
the analytically continued instantons \cite{Ai:2019fri},
which become singular in the strict real-time limit
as we mentioned above.
%% assuming that the relevant saddle points
%% in the real-time path integral
%% are given by the complex trajectories obtained by analytically continuing
%% the real solutions in the imaginary-time formalism that describe
%% quantum tunneling \cite{Ai:2019fri}.
On the other hand, 
%a particular setup
by dealing with the real-time evolution of the \emph{density matrix},
%in the Wigner representation,
%with an initial density matrix that
%it was found that
quantum tunneling can be described solely
by real classical solutions and the associated thimbles
for a positive definite initial density matrix
such as the ones given by
the Gaussian distribution \cite{Mou:2019tck,Mou:2019gyl}.
%with fluctuating initial values \cite{Mou:2019tck,Mou:2019gyl}.
Similar ideas are used also in quantum field theory
to calculate the decay rate of a false vacuum within the semi-classical
approximation \cite{Braden:2018tky,Hertzberg:2019wgx}.
%Thus it is fair to say that a complete understanding is yet to be established,
%which is precisely the aim of this paper.
%It is thus one of the goal of this paper to clarify this issue.

%In contrast to the previous works,
Here we deal with the real-time evolution of the wave function 
for a \emph{finite} time,
%in the real-time path integral,
and show that quantum tunneling in that case 
is described by 
%non-singular 
\emph{regular}
complex trajectories by explicit Monte Carlo calculations.
%which can be probed by experiments in principle using the weak measurement.
Thus we hope to provide
%establish
a new picture of quantum tunneling,
which is complementary to the one provided by the recent works mentioned above.
The physical meaning of the complex trajectories and the transition to
classical dynamics shall also be discussed.

The main obstacle in performing first-principle calculations
in the real-time path integral by using a Monte Carlo method
is
%The main obstacle is
the severe sign problem, which occurs due to the
%fact that the transition amplitude is an
%oscillatory integral over the paths with an
integrand involving
%includes
an oscillating factor $e^{iS[x(t)]}$, where the action $S[x(t)] \in \bbR$
%is real depending
depends on the path $x(t)$.
In fact, the Picard-Lefschetz theory suggests a way to overcome this problem;
namely one deforms the integration contour
numerically by the anti-holomorphic gradient flow
for a fixed amount of flow time
so that the problem becomes mild enough to be dealt with by reweighting.
This is nowadays known as the generalized
Lefschetz thimble method (GTM) \cite{Alexandru:2015sua},
which, in particular, makes the calculations possible without
prior knowledge of the relevant saddle points and the associated thimbles
unlike the earlier
proposals \cite{Fujii:2013sra,Cristoforetti:2012su,Witten:2010cx}.
%using the Lefschetz thimbles

Recently there have been further important developments of this method.
First, an efficient algorithm to generate a new configuration
was developed
based on
the Hybrid Monte Carlo algorithm (HMC), which is
applied to the variables
after the flow \cite{Fujii:2013sra,Fukuma:2019uot}
or before the flow \cite{Fujisawa:2021hxh}.
The former has an advantage that the modulus of the Jacobian
associated with the change of variables is included in the HMC procedure
of generating
a new configuration, whereas the latter has an advantage that
the HMC procedure
%of generating a new configuration
%the ficticious Hamiltonian dynamics that has to be dealt with in the HMC
simplifies drastically 
without increasing the cost
as far as one uses the backpropagation to calculate
the HMC force.
%substantially.
Second, the integration
% with respect to
of the flow time within an appropriate range
has been proposed \cite{Fukuma:2020fez}
to overcome the multi-modality problem that occurs when there are contributions
from multiple thimbles that are separated far from each other in the
configuration space. This proposal is a significant improvement over
the related ones \cite{Fukuma:2017fjq,Alexandru:2017oyw,Fukuma:2019uot}
based on tempering with respect to the flow time,
which requires the calculation of the Jacobian when one swaps the replicas.
Third, it has been realized that,
when the system size becomes large,
there is a problem that occurs
in solving the anti-holomorphic gradient flow equation,
which can be cured by optimizing the
%has been realized, and an optimal
flow equation with a kernel acting on the
drift term \cite{precondition2022}.
%% Third the problem of the anti-holomorphic gradient flow equation
%% that occurs when the system size becomes large
%% has been realized, and an optimal
%% flow equation with a kernel acting on the
%% drift term has been proposed as a remedy \cite{precondition2022}.

In this paper we apply the GTM
to the real-time path integral\footnote{See 
Refs.~\cite{Alexandru:2016gsd,Alexandru:2017lqr}
for earlier works in this direction.} for the transition amplitude
in simple quantum mechanical systems,
where the use of various new techniques mentioned above turns out to
be crucial.
%such as the one
%with a double-well potential and a quartic potential.
This, in particular,
enables us to identify the relevant complex saddle points that
contribute to the path integral from first principles, which was not
possible in the previous related works \cite{Turok:2013dfa,Tanizaki:2014xba}.
%% In particular, we find that
%% the ``weak value'' of 
%% $\hat{x}$
%% becomes complex as expected when
%% quantum tunneling
%% occurs.
By introducing a sufficiently large momentum in the initial wave function,
we find that the saddle point
%trajectories that we obtain in our simulation
%and hence their ensemble average
becomes close to real,
%and the path integral is dominated by 
which clearly indicates
%can also see
the transition to classical dynamics.

In fact, the ensemble average of the coordinate $x(t)$ at time $t$
gives the ``weak value'' \cite{Aharonov:1988xu}
of the Hermitian coordinate operator $\hat{x}$
evaluated at time $t$ with a post-selected final wave
function,
%This quantity can be also calculated by solving the Schr\"o dinger equation.
which is a physical quantity
that can be measured by experiments (``weak measurement'')
at least in principle. 
In Refs.~\cite{TANAKA2002307,Turok:2013dfa}, it was pointed out that
the complex trajectory that describes quantum tunneling
can be probed by such experiments.
We calculate the weak value of $\hat{x}$
by taking the ensemble average numerically
and reproduce the result obtained by solving the Schr\"odinger equation,
which confirms the validity of our calculations.
While the 
%weak value of $\hat{x}$ 
obtained result turns out to be complex in general,
we find that it is not always a good indicator of quantum tunneling.
For instance, the weak value can be complex in the case 
where the path integral
is dominated by more than one \emph{real} saddle points, which
typically have different \emph{complex} weights.
Similarly, we observe that the weak value can be close to real
in the case where
the path integral has contributions from more than one \emph{complex}
saddle points.
In particular, 
when the post-selected final
wave function is chosen to be the wave function
that can be obtained by time-evolving the initial wave function,
the weak value reduces to the ordinary expectation value,
which is always real 
even in the case where quantum tunneling occurs.
%even if quantum tunneling occurs.

%, which typically occurs
%% , which typically occurs when the post-selected final
%% wave function is chosen to be close to the wave function
%% that can be obtained by time-evolving the initial wave function.

We also show that the spiral shape similar
to the analytically continued
%complexified 
instantons appears in the case of a double-well potential 
when one calculates the transition amplitude
between the initial and final wave functions,
which are chosen to be
Gaussian functions centered at the two potential minima, respectively.
Furthermore the complex trajectories we obtain for a finite time turn out to
be completely regular unlike
the analytically continued instantons
obtained in the long-time limit \cite{Cherman:2014sba}.
%. This suggests that
%the aforementioned singular behaviors of the complex 
%trajectory \cite{Cherman:2014sba} is due to the long-time limit.
%obtained with a fixed boundary condition and in the long-time limit.
%raised in Ref.~
%raised in Ref.~\cite{Cherman:2014sba}.

Thus we establish a general statement that quantum tunneling
is characterized by the contribution of complex saddle points,
%and the associated thimbles,
which can be identified by using the Picard-Lefschetz theory.
In the semi-classical limit, the corresponding 
transition amplitude is suppressed
by a factor $e^{-c/\hbar}$ with $c$ being the imaginary part of the action
for the complex saddle point, which is shown to be positive in general.
% if it is a relevant one.
%This does not rely on the semi-classical approximation unlike
%the usual argument based on instantons.
%All these results are obtained
%Note that
This statement holds
not only for a double-well potential but also 
%for a quartic potential
for a quartic potential
%%, in which there is actually no potential barrier to tunnel through,
as we confirm explicitly.
%any quantum systems
%, in which there is actually no potential barrier to tunnel through
%as we confirm explicitly in the case of a quartic potential.
%% Namely, in our picture, quantum tunneling just 
%% refers to a generic process
%% that is classically prohibited.
%%
%process that is classically prohibited.
%This confirms that quantum tunneling refers to a generic
%process that is classically prohibited.
%feature of processes that are classically prohibited.
%We also find that complex trajectories are not exclusive to

The rest of this paper is organized as follows.
In section \ref{section:QT_in_real_time}
%% we discuss how we treat quantum tunneling
%% in the real-time path integral formalism.
we show that the real-time path integral can be made well-defined
by the Picard-Lefschetz theory
and use it
%discuss how 
to characterize quantum tunneling in the semi-classical limit.
In section \ref{section:review} we briefly review some previous works
in the case of a double-well potential, which will be 
important in our analysis.
In section \ref{section:complex_traj_from_montecarlo}
we show our main results 
obtained by applying the GTM to the real-time path integral.
In particular,
we identify relevant complex saddle points,
which are responsible for quantum tunneling.
%to be complex trajectories.
We also clarify the relationship to
the singular complex trajectory obtained by analytic continuation
of the instanton solution.
%% In section \ref{section:classical_limit},
%% we discuss what happens in the semi-classical limit, which is important
%% in characterizing quantum tunneling.
Section \ref{section:summary} is devoted to a summary and discussions.
In the appendix
%\ref{sec:appendix}
%\ref{section:GTM},
we explain the details of the calculation method
%the algorithm based on the GTM
used in obtaining our main results.

%\subsection*{Points to be addressed}
%\begin{itemize}
%    \item Initial and/or final wave functions are more realistic than the propagator setup
%    \item complex classical trajectories can be seen when real classical motions are not allowed
%    \item The wave functions are also important to really see the complex-valued saddle points (momentum regulator and position complexification)
%    \item Can take the classical limit to make sure that complex trajectories are something really associated to quantum tunneling
%\end{itemize}

%\section{Real-time path integral and the Picard-Lefschetz theory}
\section{Quantum tunneling in the real-time path integral}
%Quantum tunneling in real time}
\label{section:QT_in_real_time}

In this section, 
we first make the real-time path integral
well-defined using the Picard-Lefschetz theory.
Then we provide a new picture
of quantum tunneling in the real-time path integral,
which we will establish by explicit numerical calculations later.

\subsection{the real-time path integral}
\label{sec:real-time-PI}

Quantum tunneling has been conventionally described by instantons
in the imaginary-time 
formalism \cite{coleman_1985,ColemanPaperI,ColemanPaperII}.
However, in order to gain information on 
how the tunneling actually occurs,
%the quantum state during and after the tunneling, 
it is important to describe it in the real-time
path integral formalism. Another option is to
solve the Schr\"odinger equation, which however
requires the computational cost that grows exponentially with the number
of dynamical variables and hence it is not of practical use 
in many-body systems or in field theories.

In the real-time path integral formalism, the time evolution
of the wave function $\Psi(x;t)$
is described by the integral such as
\begin{equation}
    \Psi(x_\text{f};T)=\int_{x(T)=x_\text{f}} 
\!\!\!\!\!\!\!\!\!\!\!\!\!\!
{\cal D}x \,
%\Psi_\text{i}(x(0))
\Psi(x(0))
% \exp\left(iS[x(t)]\right) \ ,
\,  e^ {iS[x]/\hbar} \ ,
    \label{eq:time_evolution_path_integral}
\end{equation}
where $\Psi(x)\equiv \Psi(x;0)$ represents the initial wave function
and $S[x]$ is the action given by
\begin{equation}
    S[x]=\int_0^T dt
    \left\{\frac{1}{2} \, m \, \dot{x}^2(t)-V(x(t))\right\}
   % \  .
% - \log\Psi_\text{i}(x(0)) \ .
    \label{eq:Euclidean_action}
\end{equation}
as a functional of the path $x(t)$ with the time $0 \le t \le T$.
For later convenience, let us also introduce the ``effective action'' as
% including the initial wave function as
\begin{align}
    \Psi(x_\text{f};T)&=\int_{x(T)=x_\text{f}} 
\!\!\!\!\!\!\!\!\!\!\!\!\!\!
{\cal D}x \,
%\int_{x(T)=x_\text{f}} {\cal D}x(t)\,
%\Psi_\text{i}(x(0))
% \exp\left(iS[x(t)]\right) \ ,
 e^ {-S_{\rm eff}[x]} \ ,
    \label{eq:time_evolution_path_integral-eff} 
\\
    S_{\rm eff}[x]&= -\frac{i}{\hbar} \int_0^T dt
\left\{\frac{1}{2} \, m \, \dot{x}^2(t)-V(x(t))\right\} 
-\log\Psi(x(0))  \ .
% - \log\Psi_\text{i}(x(0)) \ .
    \label{eq:Euclidean_effective-action}
\end{align}

\subsection{the Picard-Lefschetz theory}
\label{sec:PL-theory}

Note that the expression 
\eqref{eq:time_evolution_path_integral-eff} 
%\eqref{eq:time_evolution_path_integral}
is actually a formal one
%not well defined 
since the path $x(t)$ has 
%infinitely many 
uncountably infinite degrees of freedom.
%For this reason, let us first 
Let us therefore discretize the time 
as $t = n \epsilon$ ($n=0 , \cdots , N$),
where $T = N \epsilon$, and introduce the discretized
dynamical variables $x_n = x(n \epsilon)$.
The path integral 
%\eqref{eq:time_evolution_path_integral}
\eqref{eq:time_evolution_path_integral-eff} 
%by the partition function
can then be represented as\footnote{Here and henceforth, 
we omit the overall normalization factor
for the wave function, which will not be 
important throughout this paper.}
%in the following discussions.}
\begin{equation}
%    \Psi(x_\text{f},T)=\int_{x_N=x_\text{f}} \prod dx_{n} \,
  %    \Psi(x_\text{f};T)=\int dx
%   Z = \int dx
  %\Psi_\text{i}(x(0)) = \int dx
  \Psi(x_{\rm f} ; T) = \int dx
%\Psi(x_0,0) 
%\, \delta(x_N - x_{\rm f})
% \exp\left(iS[x(t)]\right) \ ,
%\,  e^ {iS[x(t)]/\hbar} \ ,
\,  
e^ {-S_{\rm eff}(x)} \ ,
    \label{eq:time_evolution_path_integral-discrete}
\end{equation}
where $dx = \prod_{n=0}^{N-1} dx_{n}$
%\left( \prod_{n=0}^{N-1} dx_{n}\right) 
and 
%the effective action 
$S_{\rm eff}(x)$ is
a function of $x=(x_0 , \cdots , x_{N-1})$ given by\footnote{Note
that the log term in \eqref{eq:eff-action} has a branch cut.
This does not cause any problem below, however,
since in actual calculations we only need either
%$\frac{\partial S_{\rm eff}(x)}{\partial x}$
$\partial S_{\rm eff}(x)/\partial x$
or $\exp{(-S_{\rm eff}(x))}$.\label{footnote:log-action}}
\begin{equation}
    S_{\rm eff}(x)=
%-\frac{i\epsilon}{\hbar}
-\frac{i \epsilon}{\hbar} \sum_{n=0}^{N-1}\left\{
\frac{1}{2} \, m \left( \frac{x_{n+1}-x_n}{\epsilon}\right)^2
-\frac{V(x_{n+1})+V(x_n)}{2}\right\}
-\log\Psi(x_0)  \ ,
\label{eq:eff-action}
\end{equation}
where $x_N = x_{\rm f}$.

The integral
\eqref{eq:time_evolution_path_integral-discrete}
is still not well-defined
since it is not absolutely convergent.
Here we
use the Picard-Lefschetz theory \cite{Picard1897,Lefschetz1924}
to define this integral\footnote{More precisely, one
%This is equivalent to introducing 
introduces
a convergence factor by replacing $m$ by
$m \, e^{i\varepsilon}$ with $\varepsilon>0$ and 
%  taking 
take the $\varepsilon\rightarrow 0$ limit
%  since 
of the Picard-Lefschetz theory for the $\varepsilon$-deformed model,
which is equivalent to what we are doing here.
%%  has a smooth $\varepsilon \rightarrow 0$ limit.
% \cite{Feldbrugge:2022idb}.
Note that this regularization works even for an unbounded potential $V$
  unlike the Wick rotation.}.
%% The main obstacle in evaluating
%% \eqref{eq:time_evolution_path_integral},
%% for instance,
%% by Monte Carlo methods is the sign problem, which occurs
%% due to the oscillatory factor $e^ {iS[x(t)]/\hbar}$
%% in the integrand.
The idea is to apply Cauchy's theorem
and deform the integration contour
of $x=(x_0 , \cdots , x_{N-1})\in \bbR^N$ in $\bbC^N$
%into some contour of $z=(z_0 , \cdots , z_{N-1})\in \bbC^N$
by the anti-holomorphic gradient flow equation
\begin{align}
    \frac{d z_i(\sigma)}{d\sigma}
=\overline{\frac{\partial S_{\rm eff}(z(\sigma))}{\partial z_i(\sigma)}}
\label{anti-hol-grad-flow}
\end{align}
with the initial condition $z(0)=x\in\mathbb{R}^N$,
where $\sigma$ plays the role of the deformation parameter
and  $S_{\rm eff}(z)$ is the holomorphic generalization 
of \eqref{eq:eff-action}.
Note that \eqref{anti-hol-grad-flow} defines
a one-to-one map from $x\in\mathbb{R}^N$
to $z(\tau)\in \bbC^N$ for some $\tau$, 
which is referred to as the flow time.
We denote the deformed contour defined in this way
by $\mathcal{M}_\tau \subset \bbC^N$.

The important property of the anti-holomorphic gradient flow 
equation \eqref{anti-hol-grad-flow} is that
\begin{equation}
    \frac{d S_{\rm eff}(z(\sigma))}{d\sigma}
=\sum_i \frac{\partial S_{\rm eff}(z(\sigma))}{\partial z_i}
\frac{d z_i(\sigma)}{d \sigma}
=\sum_i \left|\frac{\partial S_{\rm eff}(z(\sigma))}{\partial z_i}\right|^2
%\overline{\frac{\partial S_{\rm eff}(z(\sigma))}{\partial z}}
\geq 0 \ ,
\label{prop-anti-hol-flow}
\end{equation}
which means that 
the imaginary part of the effective action
%is kept constant along the flow,
is constant along the flow,
whereas the real part 
$S_{\rm eff}(z(\sigma))$ keeps on growing with $\sigma$
unless one reaches some saddle point $z=z^\star$ defined by
\begin{align}
\frac{\partial S_{\rm eff}(z^\star)}{\partial z_i} &= 0 \ .
\label{saddle-point-equation}
\end{align}
Thus, in the $\tau\rightarrow\infty$ limit, 
the manifold $\mathcal{M}_\tau$ is decomposed into 
the so-called Lefschetz thimbles, each of which is associated with
some saddle point.
% $\mathcal{J}_a$, 
%each of which contains a saddle point $z^\star_a$.
The saddle points one obtains in this way are called 
``relevant''
%saddle points
in the Picard-Lefschetz theory.
In particular, the saddle points on the original integration contour
are always relevant.
Note also that there can be many saddle points that are not obtained
by deforming the original contour in this way, which are called
``irrelevant''.
%saddle points.
By comparing the integral over the thimble associated
with each relevant saddle point,
we can determine which saddle points have
important contributions to the original integral.

\subsection{characterization of quantum tunneling}
\label{sec:charcterization}

Having understood how to make sense of
the formal expression \eqref{eq:time_evolution_path_integral},
let us discuss how to characterize quantum tunneling in
the real-time path integral 
formalism.
% (See Ref.~\cite{Turok:2013dfa} for a pioneering work in this direction.). 
%% Our discussion can be viewed as a refinement of
%% the one in Ref.~\cite{Turok:2013dfa}, which discuss
%% quantum tunneling in the real-time path integral formalism.
%for a pioneering work in this direction. 
%here refines the one in Ref.~\cite{Turok:2013dfa} 
%%
%% By taking a different $\hbar \rightarrow 0$ limit
%% for the initial wave function in the argument,
%% we make clear the distinction between the quantum tunneling and
%% the classical motion 

%% Since 
%% the Picard-Lefschetz theory
%% forces us to complexify the path $x(t)$ 
%% in \eqref{eq:time_evolution_path_integral} to a complex path $z(t)$, 
%% we need to consider what kind of complex path $z(t)$ 
%% dominates the path integral.
For that purpose, we consider the semi-classical limit, 
which corresponds to taking the $\hbar \rightarrow 0$ limit
with the initial wave function $\Psi(x)$ assumed to have a
form\footnote{Similar discussions
  are given in Ref.~\cite{Turok:2013dfa}, where
%In Ref.~\cite{Turok:2013dfa},
  the initial wave function is assumed to have a form
  $\Psi(x)=\exp\{-\alpha(x-x_0)^2/\hbar \}$.
  However, this leads to
  a mixed boundary condition involving both $i x(0)$ and $\dot{x}(0)$,
  which does not allow a real solution
  like the one we have in \eqref{classicalEOM-and-init-momentum-constraint}.
  Thus our choice \eqref{Psi-profile} is crucial
 %Thus our setup has an advantage 
  in characterizing quantum tunneling 
  by making clear the difference from a classical motion.}
%distinction between quantum tunneling and a classical motion.}
\begin{align}
\Psi(x) &=  \psi(x) \exp \left(\frac{i \,p \, x}{\hbar} \right) \ .
\label{Psi-profile}
\end{align}
When we take the $\hbar \rightarrow 0$ limit,
we fix the profile function $\psi(x)$ and the parameter $p$
in \eqref{Psi-profile}
as well as the end point $x_{\rm f}$ and the total time $T$.
%% and assumed to have a compact support;
%% namely $\psi(x)=0$ for $x \notin \Delta$.
%% The parameter $p$ is fixed in the $\hbar \rightarrow 0$ limit
%% and plays the role of the initial momentum.

In the semi-classical limit,
the complex path $z(t)$ that dominates the path integral
in the Picard-Lefschetz theory is given by the relevant 
saddle point $z^\star$
that has the smallest ${\rm Re} S_{\rm eff}[z^\star]$.
%% dominates the path integral.
Using the continuum notation in section \ref{sec:real-time-PI},
%the path integral is dominated by some saddle point satisfying
%the discretized version of 
the saddle-point equation \eqref{saddle-point-equation} 
reduces 
in the $\hbar \rightarrow 0$ limit to
\begin{align}
0 &=  \delta  S[z(t)] + p \, \delta z(0) \\
&= \int_0^T dt \,
 \{ m \,  \dot{z}(t) \, \delta \dot{z}(t) -  V'(z(t)) \, \delta z \} 
+ p \, \delta z(0) \\
&=  \int_0^T dt \, 
 \{ - m \, \ddot{z}(t) -  V'(z(t)) \} \, \delta z(t)
+   m \Big\{  \dot{z}(T) \, \delta z(T) -  \dot{z}(0) \, \delta z(0) \Big\}
+ p \, \delta z(0) 
\label{delta-S-continuum-intermediate}
\\
&=   
\int_0^T dt \, 
 \{ - m \, \ddot{z}(t) -  V'(z(t)) \} \, \delta z(t)
+  \{  p   -  m  \,  \dot{z}(0) \} \, \delta z(0) \ ,
%% 0 &=  \frac{\delta }{\delta z(t)}  \{ S[z(t)]+p z(t) \delta(t) \} \\
%% &= - \{ m \ddot{z}(t) + V'(z(t)) \} + \{ - m \dot{z}(t) + p \} \delta(t) 
\label{delta-S-continuum}
\end{align}
where $\dot{~}$ and ${~}^{'}$ represent the derivative with respect
to $t$ and $z$, respectively,
and we have used $\delta z(T)=0$ in \eqref{delta-S-continuum-intermediate}.
%%.\footnote{In arriving at
%%\eqref{delta-S-continuum}, we have used the constraint $z(T)= x_{\rm f}$.}
Thus we obtain
\begin{align}
\label{classicalEOM-and-init-momentum}
 m \, \ddot{z}(t) &= - V'(z(t)) \ ,   \\
 m \, \dot{z}(0) &=  p  \ ,  \quad  z(T) = x_{\rm f} \ ,
\label{classicalEOM-and-init-momentum-constraint}
\end{align}
which represents the classical equation of motion
with the constraints on the initial momentum
and the final position.
%and the boundary condition $z(T)=x_{\rm f}$ at the final time.
%which is used in deriving \eqref{delta-S-continuum}.
Note that the solution becomes real if $z(0)\in \bbR$.

%According to the Picard-Lefschetz theory
%% The path integral is dominated by
%% the relevant saddle point with the
%% minimum ${\rm Re}S_{\rm eff}(z^\star)$.
%% Note that this quantity is zero for real solutions

Let us here assume\footnote{Here
%%Strictly speaking,
%%  holomorphic functions cannot have a compact support.
%%  Therefore, 
the profile function $\psi(x)$ we have in mind is,
%  for instance,
\emph{e.g.},
  a Gaussian function,
  which is well localized in some region $\Delta$ for any practical purposes.
  Alternatively, one can make a change of variable from $x_0$ to $\xi_0$
  through
%  such as
  $x_0 = \frac{1}{2} \{ (a+b) - (b-a)\tanh \xi_0 \}$ to
  impose $a \le x_0 \le b$ in the discretized
  formulation \eqref{eq:eff-action}.}
that the profile function $\psi(x)$ in \eqref{Psi-profile}
%is localized to some region $x \in \Delta$,
has a compact support $\Delta$.
%and use $\xi_0$ as the integration variable.}.
Then we consider a set of real solutions 
%$x(t)$
$z(t) \in \bbR$
%solve the classical equation of motion
%\eqref{classicalEOM-and-init-momentum}
with the initial condition $z(0)\in \Delta$, $\dot{z}(0)=p/m$
%with the initial condition $x(0)\in \Delta$, $\dot{x}(0)=p/m$
and define a domain $D \subset \bbR$ which is composed of $z(T)$.
%and define a domain $D \in \bbR$ which is composed of $x(T)$.
If $x_{\rm f} \notin D$,
there is no real solution to \eqref{classicalEOM-and-init-momentum}
satisfying the boundary condition
\eqref{classicalEOM-and-init-momentum-constraint}
with $z(0) \in \Delta$.
In that case, the path integral is dominated
by some complex solution $z^{\star}$.
The important point here is that
this solution $z^{\star}$ has to be a relevant saddle point,
which implies ${\rm Re}S_{\rm eff}[z^{\star}] \sim c/\hbar$ with $c>0$
due to the property \eqref{prop-anti-hol-flow}.
%Hence 
The transition amplitude
\eqref{eq:time_evolution_path_integral}
%\eqref{eq:time_evolution_path_integral-eff} 
is therefore suppressed by a factor $\exp(-c/\hbar)$
as expected for quantum tunneling,
whereas a classical motion that corresponds to a real
saddle point does not have this suppression factor.
%When there are many relevant complex saddle points,
In this way, we can characterize quantum tunneling
in the real-time path integral 
as the dominance of some relevant complex saddle point
based on the Picard-Lefschetz theory.\footnote{For finite $\hbar$, 
the saddle-point equation involves the profile function $\psi(x)$
in \eqref{Psi-profile}, and hence
it does not allow real solutions in the strict sense.
Furthermore, 
(almost) real solutions and complex solutions
can have comparable contributions to 
the path integral \eqref{eq:time_evolution_path_integral-eff}.
However, we can still identify the latter contribution as the effect
of quantum tunneling.
Thus the characterization of quantum tunneling is
valid 
%approximately 
beyond the semi-classical limit.}
%approximately even at finite finite $\hbar$, 
%
%% Also there are contributions from
%% the Lefschetz thimbles associated with these saddle points.
%% Thus the characterization of quantum tunneling gets somewhat 
%% obscured at finite $\hbar$.}.

%in the semi-classical limit

%% However, 
%% the effects of quantum tunneling are still
%% represented by the complex saddle points and the associated thimbles,
%% which implies that this picture of quantum tunneling 
%% is actually valid beyond the semi-classical approximation
%% adopted above to simplify our discussions.

Note also that in the strict classical limit ($\hbar \rightarrow 0$),
the transition amplitude has a support as a function of $x_{\rm f}$,
which is given by the domain $D$ defined above.
In fact, this domain $D$ shrinks to a point when
the support of the profile function $\Delta$ 
shrinks to a point ($|\Delta|\rightarrow 0$).
Thus the quantum dynamics reduces to the classical dynamics
by taking the two limits
1) $\hbar \rightarrow 0$ and 2) $|\Delta|\rightarrow 0$
in this order.
Our setup \eqref{Psi-profile} is useful here as well
since it allows us to take the two limits separately.
%making this statement,
%the support $\Delta$ shrinks to a point.
%%
%% although any setup that approximates this situation
%% realizes the transition to classical dynamics naturally.
%%
%the quantum-to-classical transition naturally.
%% The reason for the appearance of this situation for a macroscopic
%% object in Nature is given by decoherence due to the inevitable interaction
%% with the environment \cite{Schlosshauer:2019ewh}.
%See, e.g., Ref.~\cite{Konishi:2022vzz} for a recent discussion on this subject.

Let us note that the initial wave function $\Psi(x)$
in \eqref{eq:time_evolution_path_integral}
plays an important role in 
%the above picture.
determining the dominant saddle points.
%% as emphasized in Ref.~\cite{Turok:2013dfa}.
For instance, we can alternatively separate it as
%$\Psi(x_{\rm i},0)$ as
\begin{align}
\label{G-Psi}
    \Psi(x_\text{f};T)&=
\int dx_{\rm i} \, G(x_{\rm i},x_{\rm f};T) \, \Psi(x_{\rm i})  \ , \\ 
G(x_{\rm i},x_{\rm f};T) &= 
\int_{x(T)=x_\text{f}, x(0)=x_\text{i}} 
%\int_{\begin{array}{c} {\tiny x(T)=x_\text{f},}\\ \tiny{ x(0)=x_\text{i}} \end{array}} 
\!\!\!\!\!\!\!\!\!\!\!\!\!\!\!\!\!\!\!\! \!\!\! 
  {\cal D}x \,
%\Psi_\text{i}(x(0))
% \exp\left(iS[x(t)]\right) \ ,
\quad   e^ {iS[x]/\hbar} \ , 
    \label{eq:time_evolution_path_integral-propagator}
\end{align}
%% \begin{equation}
%%     \Psi(x_\text{f},T)=
%% \int dx_{\rm i} \, \Psi(x_{\rm i},0)
%% \int_{x(T)=x_\text{f}, x(0)=x_\text{i}} 
%% %\int_{\begin{array}{c} {\tiny x(T)=x_\text{f},}\\ \tiny{ x(0)=x_\text{i}} \end{array}} 
%% \!\!\!\!\!\!\!\!\!\!\!\!\!\!\!\!\!\!\!\! \!\!\! 
%%   {\cal D}x(t)\,
%% %\Psi_\text{i}(x(0))
%% % \exp\left(iS[x(t)]\right) \ ,
%% \quad   e^ {iS[x(t)]/\hbar} \ , 
%%     \label{eq:time_evolution_path_integral-propagator}
%% \end{equation}
and apply the same argument as above to the propagator 
$G(x_{\rm i},x_{\rm f};T)$
as has been done in Ref.~\cite{Tanizaki:2014xba}.
%remaining path integral with both ends fixed.
In that case, the boundary condition
\eqref{classicalEOM-and-init-momentum-constraint}
becomes
\begin{align}
z(0) &=  x_{\rm i}  \ ,  \quad  z(T) = x_{\rm f} 
\label{classicalEOM-and-init-position-constraint}
\end{align}
irrespectively of the initial wave function.
With this boundary condition,
%\eqref{classicalEOM-and-init-position-constraint}
there are always some real solutions
to the classical equation of motion \eqref{classicalEOM-and-init-momentum}
since the initial momentum can become arbitrarily large.
According to the Picard-Lefschetz theory,
this implies that there is no room for complex saddle points to be dominant
%at least
in the semi-classical limit.
Note, however, that 
the integration with respect to the real variable
$x_{\rm i}$ in \eqref{G-Psi} is highly oscillatory,
and in particular, it washes out the contributions of real solutions
with the initial momentum other than $p$ in \eqref{Psi-profile}.
This calls for another application of the Picard-Lefschetz theory,
which deforms the integration contour of $x_\text{i}$ into the complex plane.
For this reason, taking the semi-classical limit in evaluating
the propagator \eqref{eq:time_evolution_path_integral-propagator}
is not useful in evaluating the transition amplitude \eqref{G-Psi}
in the same limit.

This is in contrast to
%Note, however, that
the situation in the real-time evolution of the density matrix
%in the Wigner repesentation
\cite{Braden:2018tky,Hertzberg:2019wgx,Mou:2019gyl},
where the separation of the initial density matrix and the subsequent
real-time evolution with fixed initial data \cite{Mou:2019tck}
enables description
of quantum tunneling in terms of real classical solutions
%trajectories
and the associated
thimbles if the initial density matrix is chosen appropriately.
Thus the statement that quantum tunneling is described by
complex trajectories depends on how one formulates the problem.
%% For instance, if one takes the viewpoint of solving the Schr\"odinger
%% equation, there is no notion such as complex trajectories.
In the next section, we therefore make clear the context
in which the complex trajectories can be regarded as physical objects.

\subsection{complex trajectories as physical objects}
\label{sec:comp-traj-phys-obs}

%In this subsection, we show that
%the complex saddle points
%that appear when we deform the integration contour for the

%As we discussed in the previous subsection,
As we discussed in section \ref{sec:charcterization},
%\ref{section:QT_in_real_time},
quantum tunneling is described by complex saddle points
that appear when we deform the integration contour for the
real-time path integral based on the Picard-Lefschetz theory.
A natural question to ask here is whether such complex saddle points
are merely some mathematical notion that is useful 
in evaluating the transition amplitude or they have some physical meaning.
In fact, one can see that
the effects of the complex saddle points can be 
probed by
%considering
the ``weak value'' 
of the coordinate operator $\hat{x}$
with some post-selected wave function
as pointed out in Refs.~\cite{TANAKA2002307,Turok:2013dfa}.

Let us first recall that the weak value is defined as \cite{Aharonov:1988xu}
\begin{alignat}{3}
w(t) &= \frac{\langle \Phi | \, \hat{U}(T-t) \, \hat{x} \, 
\hat{U}(t) \, | \Psi \rangle}
{\langle \Phi | \, \hat{U}(T) \, |\Psi \rangle} \ , 
% \\ U(t) &= \exp(-it\hat{H}) \ ,
\label{def-weak-value}
\end{alignat}
where 
$\hat{U}(t) = \exp(-i \, t\hat{H}/\hbar)$ is 
the time-evolution unitary operator with the Hamiltonian $\hat{H}$.
The quantum states $|\Psi \rangle$ and $|\Phi \rangle$ correspond to the
initial wave function and the post-selected wave function, respectively.
%%which we assumed to be normalized by
%%$\langle \Psi |\Psi \rangle = \langle \Phi |\Phi \rangle = 1$.
If we choose the latter as $|\Phi \rangle = \hat{U}(T) \, |\Psi \rangle$,
the weak value $w(t)$ reduces to the usual expectation value 
\begin{alignat}{3}
w(t) &= \frac{
\langle \Psi | \, \hat{U}^\dag(t) \, \hat{x} \,  \hat{U}(t) \, 
 | \Psi \rangle}
{\langle \Psi |\Psi \rangle}  \ ,
% \\ U(t) &= \exp(-it\hat{H}) \ ,
\label{eq:usual-EV}
\end{alignat}
which implies that the weak value generalizes the notion of the expectation
value by specifying the final state $|\Phi \rangle$ to be different from 
$\hat{U}(T) \, |\Psi \rangle$.
Note that 
the weak value 
%of a Hermitian operator 
is complex in general
unlike the expectation value, which is real
for a Hermitian operator such as $\hat{x}$.
%The weak value
It is not only a mathematically well-defined quantity
but also a physical quantity that can be measured by experiments
using the so-called ``weak measurement'' \cite{Aharonov:1988xu}.

Now the crucial point for us is that
the weak value can be expressed in the real-time path integral formalism as
\begin{alignat}{3}
  \label{eq:weak-value_path-integral}
%        {eq:weak-value_path-integral}
w(t)&=\frac{1}{Z} \int
{\cal D}x \, x(t) \,
%\Psi_\text{i}(x(0))
\Psi(x(0))\, \overline{\Phi(x(T))}
% \exp\left(iS[x(t)]\right) \ ,
\,  e^ {iS[x]/\hbar} \ , \\
    Z&=\int
{\cal D}x \,
%\Psi_\text{i}(x(0))
\Psi(x(0))\, \overline{\Phi (x(T))}
% \exp\left(iS[x]\right) \ ,
\,  e^ {iS[x]/\hbar} \ ,
    \label{eq:weak-value_path-integral2}
\end{alignat}
where the action $S[x]$ is given by \eqref{eq:Euclidean_action},
and $\Psi(x)$, $\Phi(x)$ represent the wave functions
of the quantum states $|\Psi \rangle$, $|\Phi \rangle$, respectively.
%% The denominator represents the transition amplitude from
%% the initial state
%% $|\Psi \rangle$ to the final state $|\Phi \rangle$
%% up to a normalization factor,
In particular, if we choose $\Phi(x)=\delta(x-x_{\rm f})$,
the denominator $Z$ is nothing but the time-evolved wave function
\eqref{eq:time_evolution_path_integral} discussed earlier.

Note that the path integral 
\eqref{eq:weak-value_path-integral}
can also be evaluated by the Picard-Lefschetz theory.
In particular, if there is only one saddle point $z^{\star}(t)$ 
that dominates the path integral 
in the semi-classical $\hbar \rightarrow 0$ limit,
the weak value is given by $w(t) = z^{\star}(t)$.
Therefore, in order to see whether the dominant saddle point
is real or complex
in the evaluation of the time-evolved wave function
\eqref{eq:time_evolution_path_integral} at $x=x_{\rm f}$,
we just have to measure the weak value with the post-selected
wave function chosen to be $\Phi(x)=\delta(x-x_{\rm f})$.
%In order to obtain a complex weak value $w(t)$
In particular, we obtain a complex weak value $w(t)$
in the $\hbar \rightarrow 0$ limit
if the end point $x_{\rm f}$
% has to be
is located outside the domain $D$ defined
below \eqref{classicalEOM-and-init-momentum-constraint}.
%% In the case of a more general post-selected wave function $\Phi(x)$,
%% one obtains a complex weak value $w(t)$
%% in the $\hbar \rightarrow 0$ limit
%% if $\Phi(x)$ has a support outside the domain $D$.
%%
%On the other hand,
%%
%% Note, in particular, that if one uses
%% a post-selected wave function corresponding to
%% $|\Phi \rangle = \hat{U}(T) \, |\Psi \rangle$,
%% the weak value $w(t)$ is always real since it is nothing but
%% the expectation value
%% \eqref{eq:usual-EV}
%% of $\hat{x}$ at time $t$ for the initial
%% quantum state $|\Psi \rangle$.
%%
%Note, however, that the reason for this is \emph{not}
%% The reason is that the wave function
%% %post-selected wave function
%% $\Phi(x)$ in this case has a support within the domain $D$
%% in the $\hbar \rightarrow 0$ limit.

In general, the path integral is not dominated by a single saddle point,
but many saddle points can contribute comparably.
In that case, the weak value is given by a weighted average
of the saddle points $z^{\star}(t)$ with the weight
$\exp(-S_{\rm eff}[z^\star])$ being complex in general.
Therefore it is possible that the weak value becomes close to real
due to cancellation in the imaginary part
even if the path integral is dominated by some complex saddles
as we see later in section \ref{sec:post-selected-wave-function}.
For instance, if one uses
a post-selected wave function corresponding to
$|\Phi \rangle = \hat{U}(T) \, |\Psi \rangle$,
the weak value $w(t)$ is always real since it is nothing but
the expectation value
\eqref{eq:usual-EV}
of $\hat{x}$ at time $t$ for the initial
quantum state $|\Psi \rangle$.
On the contrary, it can also happen that
the weak value becomes complex 
%% even if the path integral is
%% dominated by more than one real saddles due to interference
even if the path integral is
dominated by real saddle points due to interference
as we see later in section \ref{sec:quartic}.
Thus a complex weak value is neither a necessary condition
nor a sufficient condition
for non-negligible contribution from complex saddle points in general.

Note also that unlike the expectation value, the weak value
cannot be obtained by the density matrix.
In particular, when one describes quantum tunneling using
the density matrix \cite{Braden:2018tky,Hertzberg:2019wgx,Mou:2019gyl},
one can only probe the real-time evolution of the expectation value.
Therefore, the fact that complex saddle points do not appear in the path integral
formalism for the density matrix \cite{Mou:2019tck}
with an appropriate initial condition
does not contradict the assertion here that quantum
tunneling is described by complex saddle points.

%If the semi-classical limit is not assumed,
%Beyond the semi-classical approximation,
%% There are also contributions from
%% the integration over the thimbles associated with each saddle point.
%% However, it is expected that the contribution of complex saddle points
%% can still be probed by the weak value $w(t)$.

%\subsection{Saddle points in the case of double-well potential}
\section{Brief review of previous works}
%the case with a double-well potential}
\label{section:review}

In this section
we review some previous works which will be important 
in our analysis in Section \ref{section:complex_traj_from_montecarlo}.
First we review Ref.~\cite{Tanizaki:2014xba},
%Let us here consider
%following Ref.~\cite{Tanizaki:2014xba}. 
%% The saddle points are defined by the points 
%% where the derivative of the action vanishes; 
%% i.e. the classical solutions, which will be assumed to be complex. 
in which all the solutions to the classical equation of motion
%\eqref{classicalEOM-and-init-momentum} 
%saddle points 
%including the complex ones
were obtained analytically
in the case of a double-well potential\footnote{See
Ref.~\cite{Turok:2013dfa} for earlier results
in the case of an unbounded potential with a local minimum.}
although it was not possible to identify
the relevant complex solutions
% that describe quantun tunneling.
from the viewpoint of the Picard-Lefschetz theory.
%complex saddle points.
%it was not possible to determine which complex solutions
%are responsible for quantun tunneling.
Then we review Ref.~\cite{Cherman:2014sba},
which discusses
%some discussions on this issue
the analytic continuation of the instanton solution
in the imaginary-time formalism.
%% Finally we review the proposal that the complex trajectories
%% that represent quantum tunneling can be probed 
%% by experiments 
%% using the so-called weak measurement \cite{Turok:2013dfa}.

%These results will be used later to identify the relevant saddle points
%from the configurations generated in our simulation.

\subsection{exact classical solutions for a double-well potential}
\label{sec:exact-solutions}

Let us consider a quantum system described by the action
\eqref{eq:Euclidean_action} in the continuous time formulation
with a double-well potential
\begin{equation}
    V(x)=\lambda (x^2-a^2)^2 \ ,
    \label{eq:double-well_potential}
\end{equation}
which is a typical example used to discuss quantum tunneling.
Here we take $\lambda=1/2$ and $a=1$ 
in the potential \eqref{eq:double-well_potential}
and set $m=1$ in the action \eqref{eq:Euclidean_action}
without loss of generality.

From the classical equation of motion \eqref{classicalEOM-and-init-momentum},
one can derive
%One way to write the constraint equation is 
%to consider the energy conservation
the complex version of the energy conservation
% given by 
\begin{equation}
    \left(\frac{dz}{dt}\right)^2+(z^2-1)^2=q^2 \ ,
    \label{eq:double-well_energy_conservation}
\end{equation}
where $q$ is some complex constant. 
This differential equation can be readily solved as
%This differential equation has a general solution written as
%
%% as the Jacobi elliptic function\footnote{We use the same notation 
%% for the elliptic function $\text{sd}(z,k)$ and elliptic integral $K(k)$ 
%% as in Ref.~\cite{Tanizaki:2014xba} where $k$ is the elliptic modulus.}
\begin{equation}
  z(t)=\sqrt{\frac{q^2-1}{2q}} \,
  \text{sd}
\left(\sqrt{2q}(t+c),\sqrt{\frac{1+q}{2q}}\right) \ ,
\label{zt-sol}
\end{equation}
where $c$ is another complex constant
and $\text{sd}(x,k)$ is the Jacobi elliptic function.
Thus the general solution to
\eqref{classicalEOM-and-init-momentum}
can be parametrized by
the two integration constants $q$ and $c$.

Let us fix the end points of the solution to be $z(0)=x_\text{i}$ and 
$z(T)=x_\text{f}$, which can be complex 
in general.\footnote{Note that $x_\text{i}$ and $x_\text{f}$
are assumed to be real in Ref.~\cite{Tanizaki:2014xba}
since the authors were evaluating
the propagator \eqref{eq:time_evolution_path_integral-propagator}.
As we discussed at the end of section \ref{sec:charcterization}, 
however, it is important to include the initial wave function
in the analysis, which requires us to generalize the solutions
to complex $x_\text{i}$. When we discuss the weak value
with the post-selected final wave function
as in section \ref{sec:comp-traj-phys-obs},
we have to make $x_\text{f}$ complex as well.}
%can be complex in general.
Then one finds that the parameter $k=\sqrt{(1+q)/2q}$,
which is called the elliptic modulus,
must satisfy the condition \cite{Tanizaki:2014xba}
\begin{align}
&    n\omega_1(k) +m\omega_3(k) \nonumber\\
& =\frac{T}{2}+\sqrt{\frac{2k^2-1}{8}}
   \left[\text{sd}^{-1}
   \left(\frac{\sqrt{2k^2-1}}{k\sqrt{2-2k^2}}x_\text{i},k\right)
 -(-)^{n+m}\text{sd}^{-1}
   \left(\frac{\sqrt{2k^2-1}}{k\sqrt{2-2k^2}}x_\text{f},k\right)\right]
    \label{eq:KT_constraint}
\end{align}
with $\omega_1(k)$ and $\omega_3(k)$ defined by
\begin{align}
    \omega_1(k)&=K(k)\sqrt{k^2-\frac{1}{2}} \ ,\\
    \omega_3(k)&=iK\left(\sqrt{1-k^2}\right)\sqrt{k^2-\frac{1}{2}} \ ,
\end{align}
where $K(k)$ is the complete elliptic integral of the first kind.
Note that the solution after fixing the end points still depends on 
two integers $(n,m)$, which we will refer to as ``modes'' of the solution
in what follows. 
Once we have $k$ and hence $q=1/(2k^2-1)$ for a given mode $(m,n)$, 
we can determine the complex parameter $c$ in \eqref{zt-sol}
%numerically 
from $z(0)=x_\text{i}$.

For each solution $z(t)$ obtained above,
we can obtain a solution $ \tilde{z}(t)= a z(a\sqrt{2\lambda}t)$
for arbitrary $\lambda$ and $a$ in \eqref{eq:double-well_potential}
that satisfies the boundary conditions $\tilde{z}(0)=a x_\text{i}$ 
and $\tilde{z}(T)=a x_\text{f}$.

%\subsection{Connection to the Euclidean approach}
%% The result above contains all the saddle points of the action 
%% with the double-well potential \eqref{eq:double-well_potential}. 
%% It is not possible to address which one of these are relevant 
%% with the information at hand. 

\subsection{analytic continuation of the instanton}
\label{sec:anal-cont-instanton}

Here we discuss a complex classical solution that can be obtained by
analytic continuation of the instanton solution
in the imaginary-time formalism \cite{Cherman:2014sba}.

\begin{figure}
    \centering
    \includegraphics[scale=0.67]{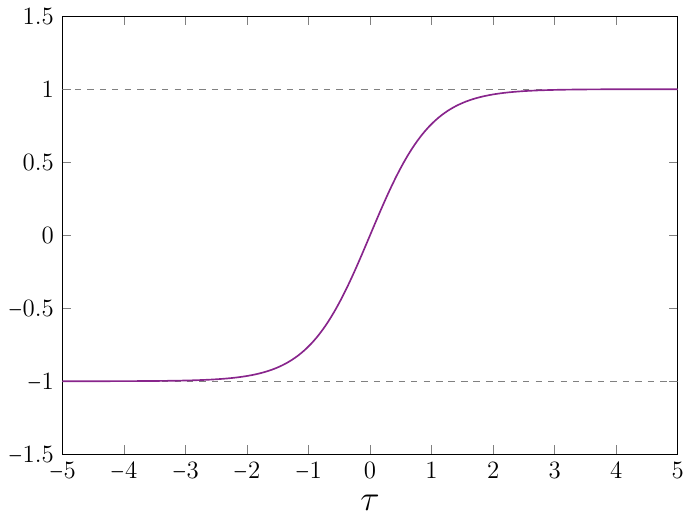}
    \includegraphics[scale=0.67]{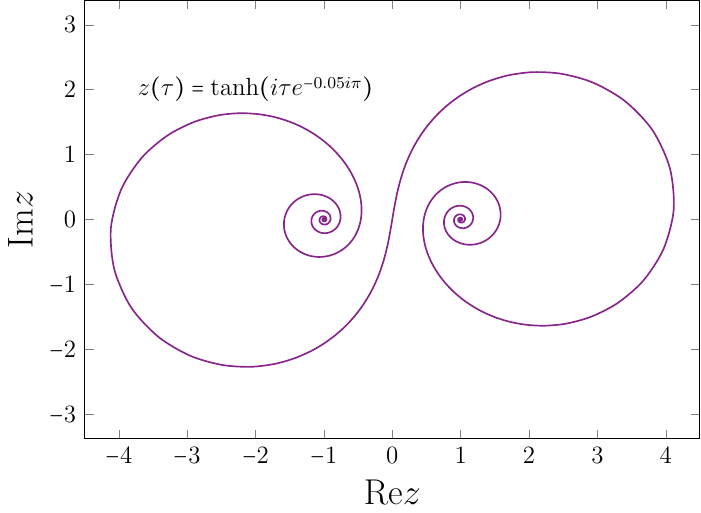}
    \caption{(Left) The instanton solution \eqref{eq:imaginary_instanton}
      in the imaginary time formalism ($\alpha=\frac{\pi}{2}$), which
      connects the two potential minima indicated by
      the horizontal dashed line.
      (Right) The trajectory of the
      complex solution
      obtained by analytic continuation of the instanton solution is
      shown in the complex plane for $\alpha=0.1 \times \frac{\pi}{2}$.}
    \label{fig:imaginary_instanton}
\end{figure}

For that, we consider the Wick rotation $t= e^{-i\alpha} \tau$,
where $\tau \in \bbR$ runs from $-\infty$ to $\infty$.
In particular, $\alpha=0$ corresponds
to the real time and $\alpha=\pi/2$ corresponds the imaginary time.
Then the action \eqref{eq:Euclidean_action} becomes
\begin{equation}
  S[x]= \frac{1}{2} \int_{-\infty}^{\infty} d\tau
\left\{ e^{i\alpha} \dot{x}^2(\tau)- e^{-i\alpha} (x^2(\tau)-1)^2 \right\} \ ,
% - \log\Psi_\text{i}(x(0)) \ .
    \label{eq:Euclidean_action-wick-rotated}
\end{equation}
where $\dot{~}$ represents
the derivative with respect to $\tau$.
The classical equation of motion reads
\begin{align}
  \ddot{z}(\tau) = - 2 e^{-2 i\alpha} z(\tau) (z^2(\tau) -1 ) \ .
  \label{complexified-classicalEOM}
\end{align}
%In the imaginary-time formaliism, which coresponds to $\alpha=\frac{\pi}{2}$,
For $\alpha=\frac{\pi}{2}$,
we obtain a real solution\footnote{In fact, the general solution
  satisfying the boundary condition \eqref{instanton-bc}
  is $z^\star(\tau)=\tanh  (\tau-\tau_0)$, where $\tau_0$ is an arbitrary
  parameter. Here we set $\tau_0=0$ since it does not affect our discussion.}
%instanton solution
\begin{equation}
%    z(\tau)=\text{tanh}(\tau-\tau_0) \ ,
  z^\star(\tau)=\tanh
  %\text{tanh} \,
  \tau \ ,
    \label{eq:imaginary_instanton}
\end{equation}
which satisfies the boundary condition
\begin{align}
  z^\star(-\infty)=-1 \ , \quad z^\star(\infty)=1
  \label{instanton-bc}
\end{align}
and therefore connects the two potential minima
as we plot in Fig.~\ref{fig:imaginary_instanton} (Left).
This is the instanton solution in the imaginary-time formalism,
and it actually describes quantum tunneling as is discussed, for instance,
in Ref.~\cite{coleman_1985}.
%$z(-\infty)=-1$ and $z(\infty)=1$,
%$\tau$ represents the imaginary time and
%where $\tau_0$ is an arbitrary constant.
%for $\tau_0=0$.

By making an analytic continuation of \eqref{eq:imaginary_instanton},
we can obtain a solution to \eqref{complexified-classicalEOM}
for arbitrary $0<\alpha \le \frac{\pi}{2}$,
which is given by 
\begin{align}
    z^\star(\tau) = \tanh \Big( i \tau  e^{-i\alpha} \Big)
%    z^\star(\tau) = \tanh \Big\{\tau  e^{-i(\alpha-\frac{\pi}{2})} \Big\}
  \label{instanton-anal-continued}
\end{align}
satisfying the same boundary condition \eqref{instanton-bc}.
Note that this solution is complex for $\alpha<\frac{\pi}{2}$
and it gives a trajectory with a spiral shape as shown in
Fig.~\ref{fig:imaginary_instanton} (Right) 
for $\alpha=0.1 \times \frac{\pi}{2}$.
For smaller and smaller $\alpha$, the trajectory winds 
more and more 
around the potential minima
as $\tau \rightarrow \pm \infty$ and it also extends
farther and farther in the complex plane.
Thus the solution that can be obtained by analytic continuation of
the instanton is actually singular in the $\alpha \rightarrow 0$ limit.

On the other hand,
if one plugs \eqref{instanton-anal-continued}
in the action \eqref{eq:Euclidean_action-wick-rotated},
one finds that the $\tau$ integration for different $\alpha$
is related to each other by just rotating the integration contour of $\tau$
in the complex plane,
which implies that
the action \eqref{eq:Euclidean_action-wick-rotated} is independent of $\alpha$
due to Cauchy's theorem.
Therefore, the transition amplitude one obtains for this solution
in the $\alpha\rightarrow 0$ limit is suppressed by
$e^{iS[z^\star]/\hbar}=e^{-S_{\rm E}/\hbar}$,
where $S_{\rm E}>0$ is the Euclidean action for the instanton solution
\eqref{eq:imaginary_instanton},
%in the imaginary-time formalism.
which implies that the transition amplitude can be correctly reproduced
by the complex saddle point obtained in this way
as far as one introduces an infinitesimal $\alpha>0$ as a kind of regulator.
In fact,
this is confirmed recently including one-loop corrections \cite{Ai:2019fri},
where the decay rate of a false vacuum
has been reproduced correctly based on the optical theorem.
%% While these results confirm that complex saddle points are indeed
%% responsible for quantum tunneling,
However,
the singular behaviors in the strict real-time limit $\alpha \rightarrow 0$
%are still somewhat worrisome.
still requires clarification.
%In particular, 
This is important, in particular, 
since complex trajectories are 
%not merely some mathematical objects that
%appear in evaluating the real-time path integral, but they are 
actually
physical objects that can be probed by experiments at least in principle
by the so-called weak measurement
as we have discussed in subsection \ref{sec:comp-traj-phys-obs}.

%\section{Complex trajectories from Monte Carlo simulations}
%\section{Results of the Lefschetz thimble calculations}
\section{Monte Carlo results obtained by the GTM}
\label{section:complex_traj_from_montecarlo}

In this section we present our results
obtained by Monte Carlo calculations using the GTM.
The partition function is given by
the transition amplitude
\eqref{eq:weak-value_path-integral2},
%with a post-selected wave function $\Phi(x)$,
which is discretized as
%which can be discretized as
\begin{equation}
%    \Psi(x_\text{f},T)=\int_{x_N=x_\text{f}} \prod dx_{n} \,
  Z   =\int dx
%\Psi_\text{i}(x(0))
%\Psi(x_0,0) 
%\, \delta(x_N - x_{\rm f})
% \exp\left(iS[x(t)]\right) \ ,
%\,  e^ {iS[x(t)]/\hbar} \ ,
\,  
e^ {-S_{\rm eff}(x)} \ ,
    \label{eq:time_evolution_path_integral-discrete-Phi}
\end{equation}
where $dx = \prod_{n=0}^{N} dx_{n}$
%\left( \prod_{n=0}^{N-1} dx_{n}\right) 
and 
\begin{equation}
    S_{\rm eff}(x)=
%-\frac{i\epsilon}{\hbar}
-\frac{i \epsilon}{\hbar} \sum_{n=0}^{N-1}\left\{
\frac{1}{2} \, m \left( \frac{x_{n+1}-x_n}{\epsilon}\right)^2
-\frac{V(x_{n+1})+V(x_n)}{2}\right\}
-\log\Psi(x_0)  -\log \overline{\Phi(x_N)}
 \ .
\label{eq:eff-action-Phi}
\end{equation}
The initial wave function \eqref{Psi-profile} is chosen as
%% 
%% where the profile function $\psi(x)$ is chosen to be
%% a Gaussian function
%that we consider is the Gaussian wave function
\begin{equation}
%  \Psi_\text{i}(x)=
  %\psi(x)=
  \Psi (x) = 
  \frac{1}{(2\pi)^{1/4}\sigma^{1/2}}
%  \sqrt{\sigma\sqrt{2\pi}}}
  %    \exp\left\{-\frac{1}{4\sigma^2}(x-b)^2+ikx\right\} \ .
  \exp\left\{-\frac{1}{4\sigma^2}(x-b)^2 + \frac{ipx}{\hbar} 
  \right\} \ .
    \label{eq:gaussian_ansatz}
\end{equation}
If we choose the post-selected wave function as
$\Phi(x)=\delta(x-x_{\rm f})$,
%in particular,
which amounts to fixing
the end point 
%is fixed to $x_N=x_{\rm f}$
to $x_N=x_{\rm f}$,
%the partition function
eq.~\eqref{eq:time_evolution_path_integral-discrete-Phi}
%gives
reduces to 
the time-evolved wave function
\eqref{eq:time_evolution_path_integral-discrete}.
%as a function of $x_{\rm f}$.
%
%where $\sigma$ is the width, $b$ is the center,
%and $k$ is the average wave number of the initial wave function.
%The end point is fixed to $x_{\rm f}=1$ throughout this section.

In all the simulations in this work,
we set the mass to $m=1$
%and In actual simulation, we
and the total time to $T=2$,
which is
%by choosing the scale of length and time appropriately.
%The time coordinate is then
divided into $N=20$ intervals.
%which implies that the lattice spacing for time is $\epsilon=0.1$.
Except in subsection \ref{section:classical_limit},
where we discuss the semi-classical $\hbar \rightarrow 0$ limit,
we set $\hbar=1$.
See Appendix \ref{sec:appendix}
%\ref{section:GTM}
for the details of the method
used for our simulations.

\subsection{the case of a double-well potential}
\label{sec:DW-pot}

In this section we consider the case of a double-well potential
\eqref{eq:double-well_potential} with $a=1$.
%and $a=1$.
%The height of the potential at the local maximum $x=0$ is $V_0= \lambda a^4$.
The height of the potential at the local maximum $x=0$ is $V_0=\lambda$.
We use $b=-1$ and $\sigma=0.3$
for the initial wave function \eqref{eq:gaussian_ansatz}
so that it is well localized around $x=-1$, which is one of the potential minima.
% $x=\pm 1$.
We use $\Phi(x)=\delta(x-x_{\rm f})$ for the post-selected wave function,
where $x_{\rm f}=1$ is chosen to be the other potential minimum.

\begin{figure}
    \centering
    \includegraphics[scale=0.8]{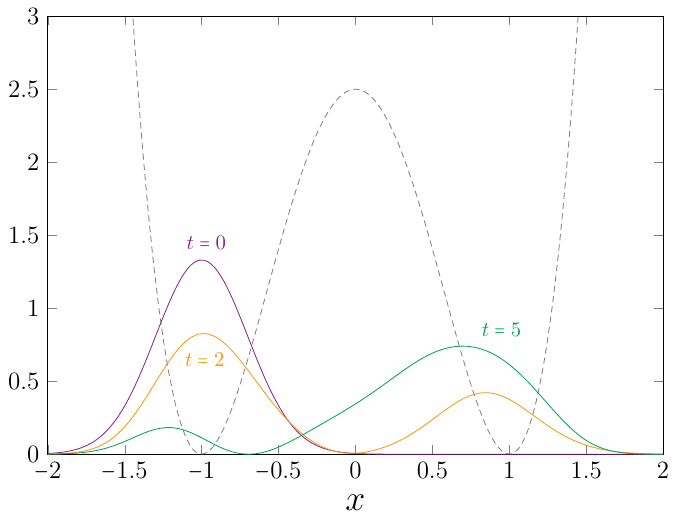}
    \caption{The distribution $|\Psi(x;t)|^2$
      at time $t=0$ (purple line), $t=2$ (yellow line)
      and $t=5$ (green line)
      are shown for the initial wave function \eqref{eq:gaussian_ansatz}
      with $\sigma=0.3$, $b=-1$, $p=0$
      in the double-well potential \eqref{eq:double-well_potential}
      with $\lambda=2.5$, $a=1$ (gray, dashed line).}
    \label{fig:example_Pt}
\end{figure}

In order to choose an appropriate value for $\lambda$ in the potential
%and $\sigma$ in the initial wave function
%appropriate parameters
to probe quantum tunneling,
we consider the probability 
\begin{equation}
    P =\sum_{E \geq V_0} |\langle E|\Psi\rangle|^2 
    \label{eq:classical_tunneling_probability}
\end{equation}
of the initial quantum state $|\Psi \rangle$
having energy larger than the
potential barrier $V_0$, where $| E \rangle$ represents the
normalized energy eigenstate with the energy $E$.
If we choose the momentum $p=0$ in
the initial wave function \eqref{eq:gaussian_ansatz},
we obtain $P \sim 0.11$ for $\lambda=2.5$.
We therefore use $\lambda=2.5$ in our calculation.\footnote{In Ref.~\cite{Mou:2019gyl},
  thimble calculations for the density matrix time-evolution were performed
  with a double-well potential $V=\frac{1}{2}\phi^2 (1-g\phi)^2$,
  which corresponds to ours \eqref{eq:double-well_potential} 
  through $x= 2g(\phi-\frac{1}{2g})$, $a=1$ and $\lambda=\frac{1}{32g^2}$.
  Their choice $g=0.3$ and $g=0.5$ for simulations
  corresponds to $\lambda \sim 0.35$
  and $0.125$, respectively, and their initial wave function 
  corresponds to choosing 
  $b=0$, $\sigma=\frac{1}{\sqrt{2}} \sim 0.71$ and $p=0$
  in \eqref{eq:gaussian_ansatz}.
  The probability \eqref{eq:classical_tunneling_probability} is given by
  $P\sim 0.54$ and $P\sim 1.0$ for $g=0.3$ and $g=0.5$, respectively.
  %% , which suggests that the process being studied correspond
  %% mostly to classical motions.
  It would be interesting to see whether their method works even in the
  case that corresponds to smaller $P$.}

\begin{figure}
    \centering
    \includegraphics[scale=0.8]{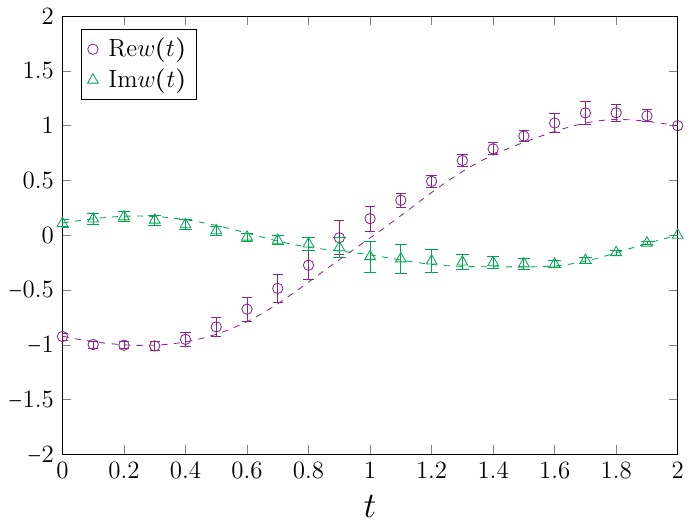}
    \includegraphics[scale=0.8]{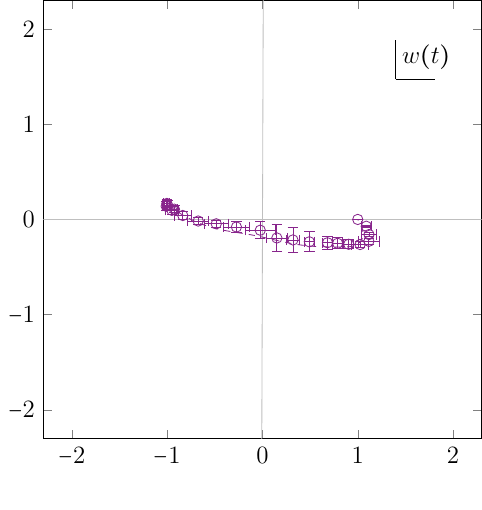}
%    \hspace{1cm}
    \includegraphics[scale=0.8]{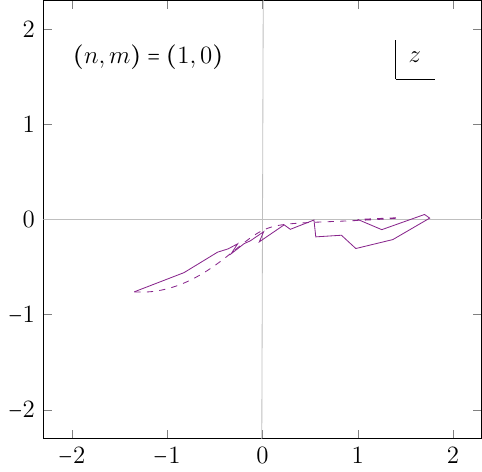}
    \includegraphics[scale=0.8]{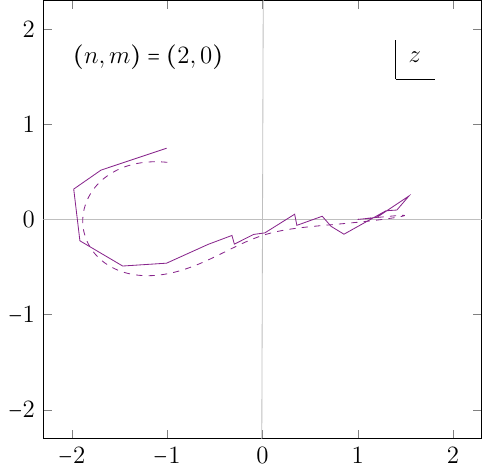}
    \caption{The results obtained
      for the initial wave function \eqref{eq:gaussian_ansatz}
      with $\sigma=0.3$, $b=-1$, $p=0$
      and the post-selected wave function $\Phi(x)=\delta(x-x_{\rm f})$
      with $x_{\rm f}=1$ 
      in the double-well potential \eqref{eq:double-well_potential}
      with $\lambda=2.5$, $a=1$.
      (Top) The weak value of the coordinate is plotted against time $t$
      in the Left panel, while the trajectory of the weak value is plotted 
      in the complex plane in the Right panel.
      The dashed lines represent the result obtained
      from \eqref{def-weak-value} by solving the Schr\"odinger equation.
(Bottom) Two typical trajectories obtained from the numerical simulation
      with the same parameters as in the Top panels.
      %for the same parameters as in the plots in the Top panels.
   The dashed lines represent the closest classical solutions 
   obtained by choosing the mode $(n,m)$ and the initial point $x_{\rm i}$
   with the final point $x_{\rm f}=1$ fixed.
%   shown on the top-left corner.
}
    \label{fig:ctrl_WM}
\end{figure}

%% \begin{figure}
%%     \centering
%%     \includegraphics[scale=0.8]{figures/numerical_results/ctrl_traj1.pdf}
%%     \includegraphics[scale=0.8]{figures/numerical_results/ctrl_traj2.pdf}
%%     \caption{Two of the typical trajectories obtained from the numerical simulation in the control setup. The dashed lines are the closest classical solutions (saddle points) to the trajectories, with the modes $(n,m)$ shown on the top-left corner.}
%%     \label{fig:ctrl_traj}
%% \end{figure}

\begin{figure}
    \centering
    \includegraphics[scale=0.8]{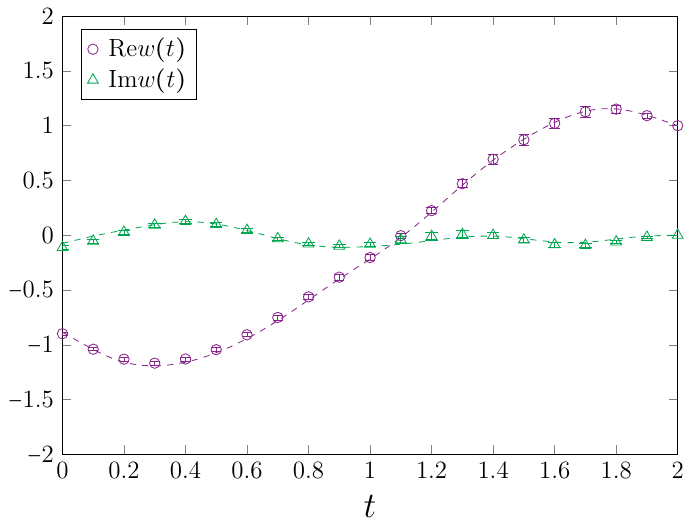}
    \includegraphics[scale=0.8]{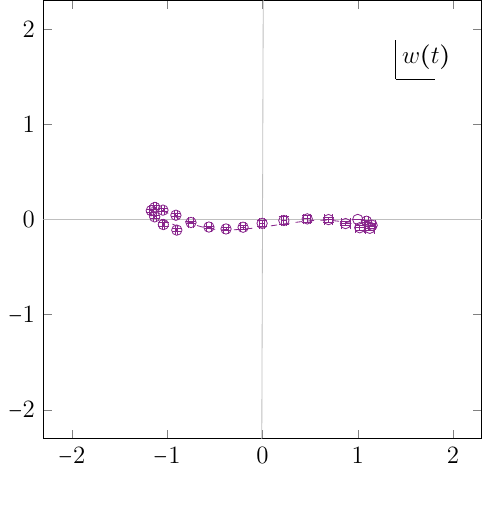}
    \includegraphics[scale=0.8]{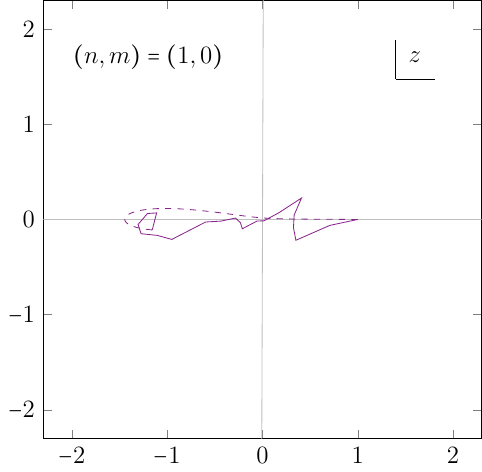}
    \includegraphics[scale=0.8]{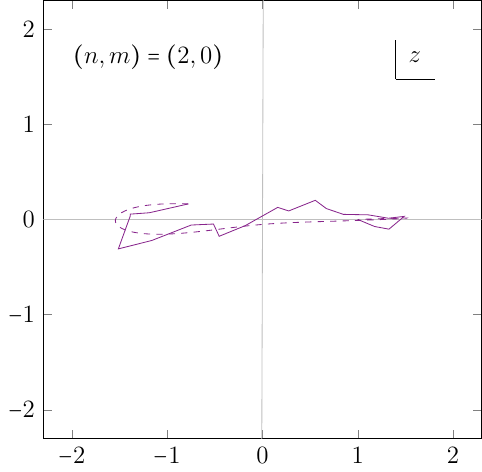}
    \caption{The results obtained
      for the initial wave function \eqref{eq:gaussian_ansatz}
      with the momentum $p=-2$. All the other parameters are the same
      as in Fig.~\ref{fig:ctrl_WM}.
      (Top) The weak value of the coordinate is plotted against time $t$
      in the Left panel, while the trajectory of the weak value is plotted 
      in the complex plane in the Right panel.
      The dashed lines represent the result obtained
      from \eqref{def-weak-value} by solving the Schr\"odinger equation.
(Bottom) Two typical trajectories obtained from the numerical simulation
   with the same parameters as in the Top panels.
   The dashed lines represent the closest classical solutions 
   obtained by choosing the mode $(n,m)$ and the initial point $x_{\rm i}$
   with the final point $x_{\rm f}=1$ fixed.}
    \label{fig:k2_WM}
\end{figure}

Note that a typical tunneling time can be evaluated by
\begin{equation}
    t_0 \sim\frac{\pi\hbar}{\Delta E} \ ,
\end{equation}
where $\Delta E$ is the energy difference between
the ground state and the first excited state.
For $\lambda=2.5$, we find $t_0 \sim 5$.
In Fig.~\ref{fig:example_Pt} we plot the wave functions
at $t=0$, $t=2$ and $t=5$
%time-evolved wave function at $T=5$ obtained
obtained for this setup 
by solving the Schr\"odinger equation 
with Hamiltonian diagonalization.
The result for $t=2$ shows that the significant portion of the distribution
has moved to the other potential minimum $x=1$,
which implies that quantum tunneling has indeed occurred.
%and $\sigma=0.3$ in what follows.

In Fig.~\ref{fig:ctrl_WM} (Top), we show our results for
the weak value $w(t)$ of the coordinate at time $t$ 
defined by \eqref{eq:weak-value_path-integral}
for the initial wave function \eqref{eq:gaussian_ansatz}
with $\sigma=0.3$, $b=-1$, $p=0$
and the post-selected wave function $\Phi(x)=\delta(x-x_{\rm f})$
with $x_{\rm f}=1$
     in the double-well potential \eqref{eq:double-well_potential}
      with $\lambda=2.5$, $a=1$.
      The dashed lines represent the results obtained 
%      by evaluating 
      directly from \eqref{def-weak-value}
% with $\Phi(x)=\delta(x-x_{\rm f})$
      by solving the Schr\"odinger equation with Hamiltonian diagonalization.
      The agreement between our data and the direct results 
confirms the validity of our calculation.
We find that the weak value $w(t)$ is indeed complex 
except for the end point, which is fixed to $w(T)=x_{\rm f}=1$.
Note, in particular, that $w(0)$ is also complex although it is
close to $x=-1$, which is the center of 
the Gaussian wave function \eqref{eq:gaussian_ansatz}.

In the Bottom panels of Fig.~\ref{fig:ctrl_WM},
 we show two typical trajectories
obtained from the simulation with the same parameters as 
in the Top panels.
These trajectories are obtained for a sufficiently long
flow time $\tau \sim 4$ in the 
GTM (See Appendix \ref{section:integrating-flow-time}.) so that
%and therefore 
they are expected to
be close to some relevant saddle points
except for fluctuations along the thimble,
which are seen as small wiggles in the observed trajectories.
Indeed we are able to find a classical solution discussed in
section \ref{sec:exact-solutions} which is close
to each of these trajectories
by choosing the mode $(n,m)$ and the initial point $x_{\rm i}$
with the final point $x_{\rm f}=1$ fixed.
%Note also that the
We find that the typical trajectories have a larger imaginary part
on the left and a smaller imaginary part on the right,
which suggests that quantum tunneling occurs first and then 
some classical motion follows.
%mostly in the first part
This feature is obscured in
%Note that this is different from the behaviors of 
the weak value $w(t)$ 
%% Note also that the typical trajectories have a larger imaginary part
%% on the left and a smaller imaginary part on the right,
%% which is different from the behaviors of the weak value $w(t)$ 
shown in the Top-Right panel.
%This is due to the fact that the weak value $w(t)$ 
This is possible since the weak value $w(t)$ 
is a weighted average
of $x(t)$ obtained from the simulation, where the 
weight \eqref{eq:reweithting-factor} is complex
in general since it consists of the phase factor $e^{-i \, {\rm Im}S_{\rm eff}}$
and the Jacobian for the change of 
variables.
% (See eq.~\eqref{eq:reweithting-factor}.).

%\subsection{introducing momentum in the initial wave function}

%In this section,
Next we introduce nonzero momentum
%in the initial wave function.
$p=-2$ in the initial wave function \eqref{eq:gaussian_ansatz}.
%in the initial wave function.
%
%$p$ in the initial wave function \eqref{eq:gaussian_ansatz}
%%
%% to see what happens when a classical motion over the potential barrier
%% becomes possible.
In Fig.~\ref{fig:k2_WM} we show our results
%for $p=-2$ in the initial wave function \eqref{eq:gaussian_ansatz}
with all the other parameters the same as in Fig.~\ref{fig:ctrl_WM}.
Since the initial kinetic energy is $p^2/2 = 2$, which is close to
the potential barrier $\lambda = 2.5$, 
%we expect the dominance of real saddle points.
a classical motion over the potential barrier is
possible if the initial point $x(0)$ is slightly shifted
from the potential minimum.
Indeed we find that the weak value
% close to real,
and the typical trajectories
% obtained by the simulation 
become close to real.

%\subsection{Connection to the complexified instanton}
\subsection{the case with a Gaussian post-selected wave function}
\label{sec:post-selected-wave-function}

\begin{figure}
    \centering
    \includegraphics[scale=0.8]{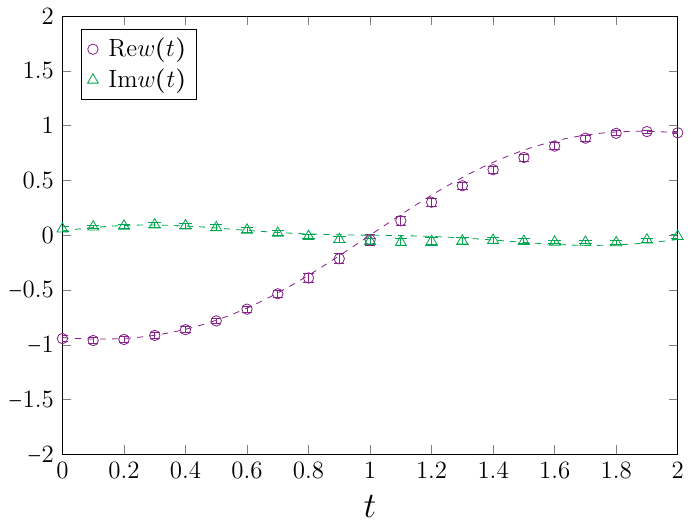}
    \includegraphics[scale=0.8]{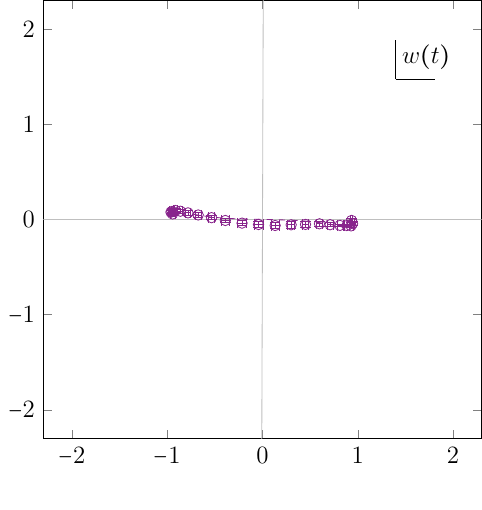}
    \includegraphics[scale=0.8]{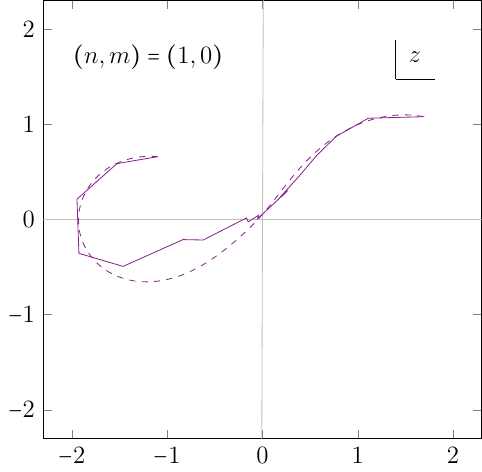}
    \includegraphics[scale=0.8]{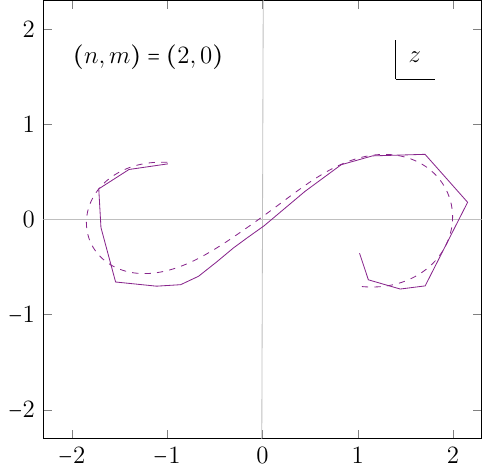}
    \caption{The results obtained for a post-selected wave function
      \eqref{parity-Phi-Psi} with 
      all the other parameters the same
      as in Fig.~\ref{fig:ctrl_WM}.
      (Top) The weak value of the coordinate is plotted against time $t$
      in the Left panel, while the trajectory of the weak value is plotted 
      in the complex plane in the Right panel.
      The dashed lines represent the result obtained
      from \eqref{def-weak-value} by solving the Schr\"odinger equation.
(Bottom) Two typical trajectories obtained from the numerical simulation
   with the same parameters as in the Top panels.
   The dashed lines represent the closest classical solutions 
   obtained by choosing not only the mode $(n,m)$
   and the initial point $x_{\rm i}$
   but also the final point
   $x_{\rm f}$ unlike the previous cases.}
   %$x_{\rm f}=1$ fixed unlike the previous cases.}
    \label{fig:fnwf_WM}
\end{figure}

So far, we have been fixing the end point to 
the other potential minimum $x_{\rm f}=1$.
This, in particular, allows us to see what kind of trajectories
dominate the real-time path integral
for the time-evolved wave function \eqref{eq:time_evolution_path_integral}
in the Picard-Lefschetz theory.
% for $x_{\rm f}=1$.
We were able to see
%% In particular, we have seen
that complex saddle points
indeed dominate
by choosing the parameters in the initial wave function and
the double-well potential appropriately.
%% for $p=0$, while real saddle points dominate for $p=-2$
%% for the other parameters for the initial wave function and
%% the double-well potential chosen appropriately.

On the other hand,
the analytic continuation of the instanton solution
suggests that
the relevant complex trajectory that describes quantum tunneling
in the case of a double-well potential
has a spiral shape shown in Fig.~\ref{fig:imaginary_instanton} (Right).
In order to clarify the relationship to this result,
we
%it is more natural to 
consider 
a post-selected wave function $\Phi(x)$ other than 
$\Phi(x)=\delta(x-x_{\rm f})$
in the real-time path integral
\eqref{eq:weak-value_path-integral2}.
%with 
%$w(t)$ defined \eqref{eq:weak-value_path-integral2} with

In fact, considering the parity symmetry $x \mapsto -x$
of the quantum system at hand,
it is natural to choose the post-selected wave function as
\begin{equation}
  \Phi(x)=\Psi(-x)  \ ,
  \label{parity-Phi-Psi}
\end{equation}
where the initial wave function $\Psi(x)$
is given by \eqref{eq:gaussian_ansatz}.
This makes the saddle-point equation \eqref{saddle-point-equation}
invariant under simultaneous reflection of
time $t \mapsto T-t$ and space $z \mapsto -z$,
and hence allows a solution with the symmetry $z(T-t)= -z(t)$,
which is compatible with the spiral shape
in Fig.~\ref{fig:imaginary_instanton} (Right).

In Fig.~\ref{fig:fnwf_WM} we show our results
for the initial wave function \eqref{eq:gaussian_ansatz}
with all the parameters the same as in Fig.~\ref{fig:ctrl_WM}
and the post-selected wave function now given by \eqref{parity-Phi-Psi}.
Unlike the previous cases with $\Phi(x)=\delta(x-x_{\rm f})$,
the end point of the trajectories is not constrained to $x(N)=x_{\rm f}$
and it flows into the complex plane due to the flow
equation \eqref{anti-hol-grad-flow}.
%{prop-anti-hol-flow}.
In particular, the typical trajectory shown
in Fig.~\ref{fig:fnwf_WM} (Bottom-Right)
has a spiral shape with the symmetry $z(T-t)= -z(t)$,
which
%looks qualitatively similar to
resembles
the trajectory
in Fig.~\ref{fig:imaginary_instanton} (Right)
obtained by analytic continuation of the instanton solution.
Furthermore the classical solutions
%that are close to the trajectories
that appear in our simulation
are all regular 
even though we are working in the strict real-time
limit $\alpha\rightarrow 0$
discussed in section \ref{sec:anal-cont-instanton}.
%unlike what we saw in section \ref{sec:anal-cont-instanton}.
It is conceivable that the spiral winds more and more
around the potential minima as we increase the time $T$.
We also note that the weak value shown in the Top-Right panel
turns out to be quite close to real,
which is possible since there are more than one
relevant complex saddle points that interfere with each other.
%one another.
We consider that the situation is similar to the case with
the post-selected quantum 
state $|\Phi \rangle = \hat{U}(T) \, |\Psi \rangle$
discussed at the end of section \ref{sec:comp-traj-phys-obs}.

\subsection{the case of a quartic potential}
\label{sec:quartic}

In this section we discuss the case of a quartic potential
\begin{equation}
  V(x)=\kappa \, x^4 \ ,
  \label{eq:quartic-pot}
\end{equation}
in which there is no potential barrier to tunnel through.
%where we choose $\kappa=1$ in what follows.
%which does not have a potential barrier to tunnel through.

\begin{figure}
    \centering
    \includegraphics[scale=0.8]{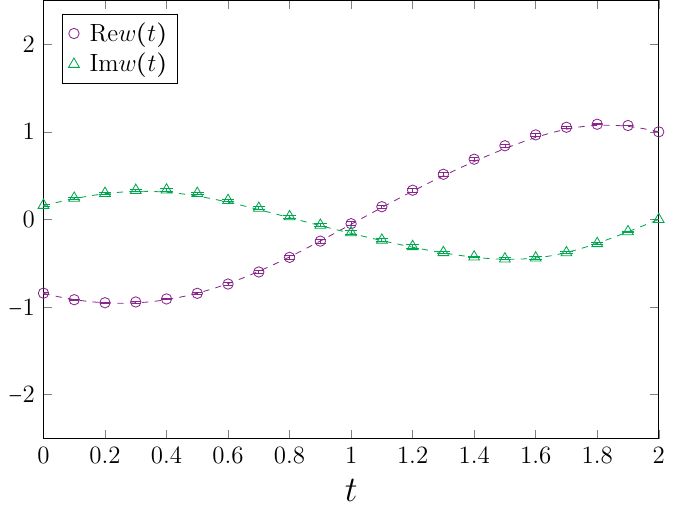}
    \includegraphics[scale=0.8]{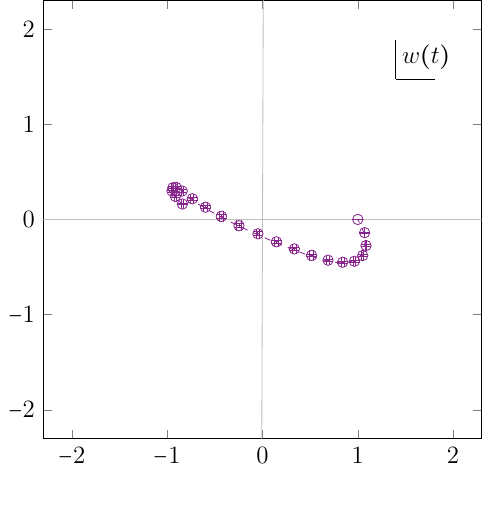}
    \includegraphics[scale=0.8]{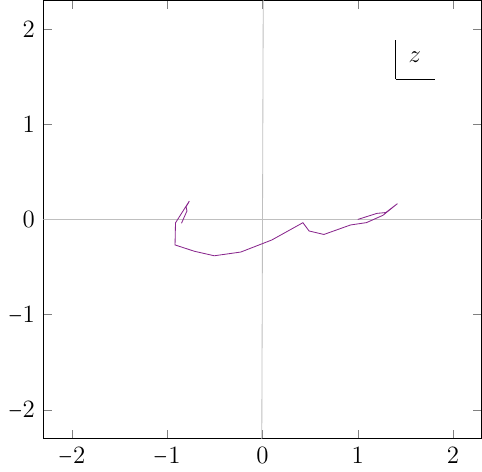}
    \includegraphics[scale=0.8]{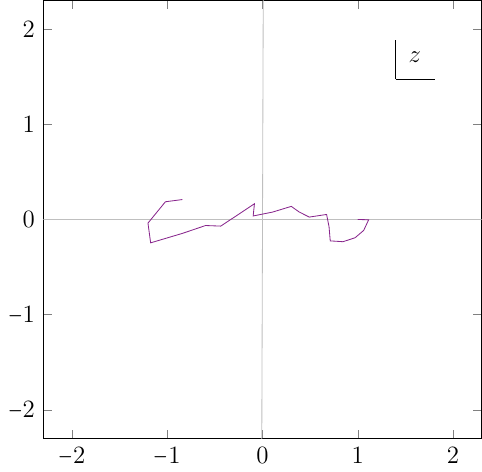}
    \caption{The results obtained for the quartic potential
      \eqref{eq:quartic-pot} with $\kappa=1$. As in Fig.~\ref{fig:ctrl_WM},
      the initial wave function is chosen as \eqref{eq:gaussian_ansatz}
      with $\sigma=0.3$, $b=-1$, $p=0$
      and the post-selected wave function is chosen as
      $\Phi(x)=\delta(x-x_{\rm f})$ with $x_{\rm f}=1$. 
      (Top) The weak value of the coordinate is plotted against time $t$
      in the Left panel, while the trajectory of the weak value is plotted 
      in the complex plane in the Right panel.
      The dashed lines represent the result obtained
      from \eqref{def-weak-value} by solving the Schr\"odinger equation.
(Bottom) Two typical trajectories obtained from the numerical simulation
   with the same parameters as in the Top panels.}
%      are shown.}
      %for the same parameters as in the plots in the Top panels.
    \label{fig:qrtc_WM}
\end{figure}

\begin{figure}
    \centering
    \includegraphics[scale=0.9]{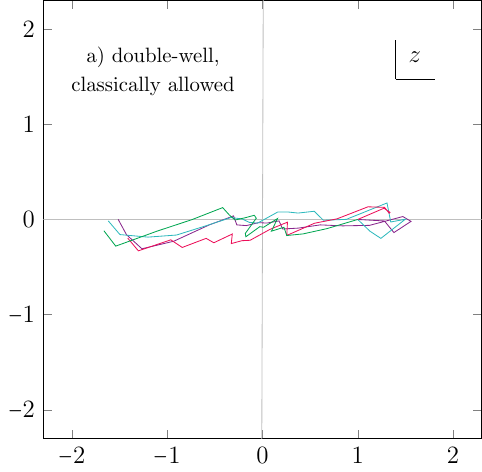}
    \includegraphics[scale=0.9]{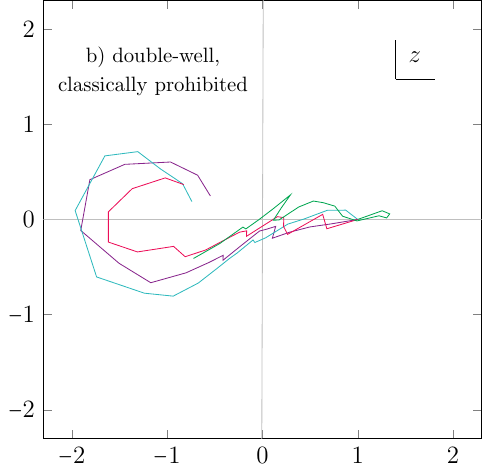}
    \includegraphics[scale=0.9]{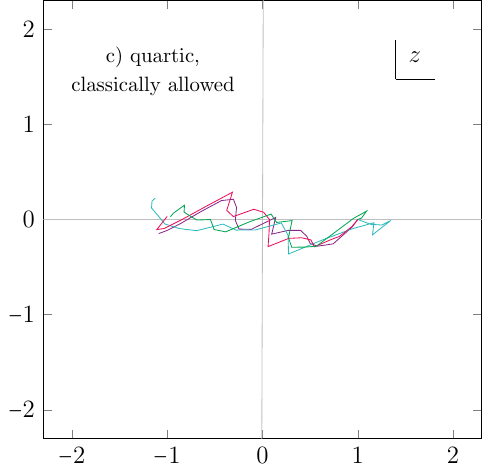}
    \includegraphics[scale=0.9]{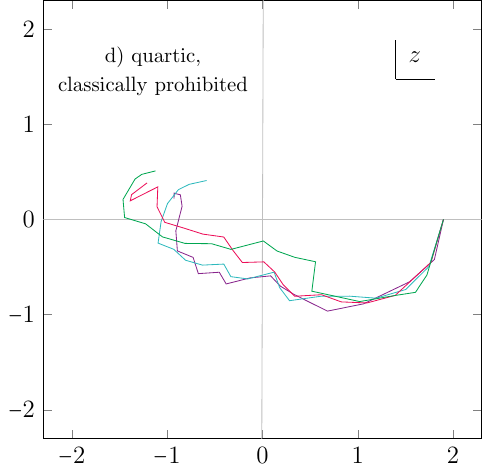}
    \caption{Typical trajectories for $\hbar=0.5$ obtained in various cases.
      (Top-Left) The case with a double-well potential
      with all the parameters other than $\hbar$ being
      the same as in Fig.~\ref{fig:ctrl_WM}.
      In particular,
      the initial wave function is chosen as \eqref{eq:gaussian_ansatz} with
      $b=-1$, $\sigma=0.3$, $p=0$.
      (Top-Right) The same as the Top-Left panel except that
      the initial wave function is chosen now as \eqref{eq:gaussian_ansatz} with
      $b=-0.8$, $\sigma=0.2$, $p=0$.
      (Bottom-Left) The case with a quartic potential
      with all the parameters other than $\hbar$ being
      the same as in Fig.~\ref{fig:qrtc_WM}.
      In particular, the end point is fixed to $x_{\rm f}=1$.
      (Bottom-Right) The same as the Bottom-Left panel except that
      the end point is now fixed to $x_{\rm f}=1.9$.}
    \label{fig:classical_limit}
\end{figure}

In Fig.~\ref{fig:qrtc_WM} we show our results for $\kappa=1$.
As in section \ref{sec:DW-pot},
the initial wave function is chosen as \eqref{eq:gaussian_ansatz}
with $\sigma=0.3$, $b=-1$, $p=0$
and the post-selected wave function
is chosen as $\Phi(x)=\delta(x-x_{\rm f})$ with $x_{\rm f}=1$.

The typical trajectories shown in the Bottom panel are close to real,
which suggests that classical motions are allowed with the chosen parameters.
Note, however, that, the weak value shown in the Top-Right panel
turns out to be complex, which can be understood as a result of
interference among various trajectories with relative complex weights.
This is in contrast to the situation
with the $p=-2$ case discussed in section \ref{sec:DW-pot},
where the typical trajectories and the weak value are both close to real.

\subsection{the semi-classical limit}
\label{section:classical_limit}

So far, we have chosen $\hbar=1$
in 
%the partition function 
\eqref{eq:eff-action-Phi}
and 
%the wave function 
\eqref{eq:gaussian_ansatz}.
In this section, we reduce it to $\hbar=0.5$
%discuss what happens if we reduce $\hbar$.
and discuss
%to see 
%This, in particular, suggests 
what happens in the semi-classical limit.
%% which plays a crucial role in characterizing quantum tunneling
%% as we discussed in section \ref{sec:charcterization}.

Let us consider the double-well potential case with $p=0$
%the $p=0$ case
shown in Fig.~\ref{fig:ctrl_WM},
%discussed in section \ref{sec:DW-pot}.
where typical trajectories are complex
for $\hbar=1$.
With all the parameters being the same, here we reduce
$\hbar$ to $\hbar=0.5$.
In Fig.~\ref{fig:classical_limit} (Top-Left), we find
that typical trajectories are close to real suggesting
the dominance of real saddle points corresponding to some classical motions.
%due to suppression of quantum effects.
These classical motions are possible
since for the initial position $x\leq-\sqrt{2}$,
the potential energy \eqref{eq:double-well_potential} of the particle
becomes larger than the potential barrier $V_0 = \lambda$.
Indeed we find that the initial point $z(0)$ is 
close to $x=-\sqrt{2}$.
%the region $x\leq -\sqrt{2}$.
Namely, for this setup, we expect that real saddle points dominate
in the $\hbar \rightarrow 0$ limit.

Here we change the parameters in
the initial wave function \eqref{eq:gaussian_ansatz}
from the previous ones $b=-1$, $\sigma=0.3$
to the new ones $b=-0.8$, $\sigma=0.2$ with $p=0$ unchanged
so that the initial wave function is almost zero
%within the region $x\leq-\sqrt{2}$.
at $x\leq-\sqrt{2}$.
In Fig.~\ref{fig:classical_limit} (Top-Right),
we indeed find that typical trajectories in this case
are complex for $\hbar=0.5$. 
Namely, for this setup, we expect that complex saddle points dominate
in the $\hbar \rightarrow 0$ limit.

Next we consider the quartic potential case shown in Fig.~\ref{fig:qrtc_WM},
%section \ref{sec:quartic}.
where typical trajectories are close to real for $\hbar=1$.
With all the parameters being the same, here we reduce
$\hbar$ to $\hbar=0.5$.
In Fig.~\ref{fig:classical_limit} (Bottom-Left), we find
that typical trajectories are still close to real suggesting
the dominance of real saddle points corresponding to some classical motions.
These classical motions are possible
since for the initial position $x \le  -1$,
the potential energy
%\eqref{eq:double-well_potential} of the particle
is larger than the potential energy $V_0 = \kappa$ at the end point $x_{\rm f}=1$.
Namely, for this setup, we expect that real saddle points dominate
in the $\hbar \rightarrow 0$ limit.

Here we change the end point from the previous one $x_{\rm f}=1$
to the new one $x_{\rm f}=1.9$
with all the other parameters unchanged.
In order to have a classical motion,
the initial position should be $x\leq - 1.9$,
where the initial wave function is almost zero for $b=-1$ and $\sigma=0.3$.
In Fig.~\ref{fig:classical_limit} (Bottom-Right),
we indeed find that typical trajectories in this case
are complex for $\hbar=0.5$. 
Namely, for this setup, we expect that complex saddle points dominate
in the $\hbar \rightarrow 0$ limit.
%% This case highlights the importance of the post-selection
%% in characterizing quantum tunneling in the semi-classical limit.
This case 
highlights
%demonstrates 
the role played by 
the post-selection in characterizing
quantum tunneling.

\section{Summary and discussions}
\label{section:summary}
In this paper we have investigated the description of quantum tunneling
in the real-time path integral formalism, in which complex trajectories
were expected to play a crucial role.
In particular, we were able to determine, for the first time,
the complex saddle points
that are relevant from the viewpoint of the Picard-Lefschetz theory
using Monte Carlo methods.
The severe sign problem that occurs in evaluating the oscillating integral
was overcome by the GTM with various new techniques developed recently.
Our results establish a statement
that quantum tunneling is characterized
by complex saddle points,
% and the associated Lefschetz thimbles,
which dominate the path integral in the semi-classical limit
when the classical motion is not allowed by boundary conditions.
%% In this characterization, the idea of post-selection plays an
%% important role. In particular, the complex trajectories can be probed
%% by looking at the weak value of the coordinate operator at intermediate time.
We have also clarified the relationship to the instanton,
which is widely used as a standard description of quantum tunneling
based on the imaginary-time path integral formalism.
%% We find, in particular, that the complex trajectories
%% obtained by analytic continuation of instantons appear only in the
%% long-time limit.
%%
%% We would like to emphasize that the real-time path integral can,
%% in general, have comparable contributions
%% from real saddle points and complex saddle points.
%% It is intriguing that the description discussed here 

%the puzzle concerning the analytic continuation
%of the instanton solution in the imaginary-time path integral.
%In particular,

Among various applications of the description of
quantum tunneling
in the real-time path integral formalism,
%it is interesting to investigate
we consider that
the false vacuum decay is important in the context
of cosmology and particle physics \cite{Hayashi:2021kro}.
%
%, has been investigated recently
%based on the Picard-Lefschetz theory \cite{Hayashi:2021kro}.
We would also like to recall that
%we recall that
quantum tunneling is expected to have 
%occurred
taken place
%is expected to be important
%also in understanding
at the beginning of our
universe \cite{Vilenkin:1982de,Hartle:1983ai}.
%In particular, 
The problem here was that there seemed to be no guiding principle
to choose the integration contour 
in the path integral formalism and hence it was not possible
to determine the saddle points that actually 
contribute \cite{Halliwell:1988ik}.
This problem was solved recently by the recognition that
quantum gravity should be formulated
using the real-time (or Lorentzian) path integral
based on the Picard-Lefschetz theory \cite{Feldbrugge:2017kzv}.
%Based on this perspective,
%In view of this perspective,
From this point of view, the quantum tunneling at the beginning
of the Universe is described by the emergence of Euclidean geometry
as a dominant complex saddle point.
% in the Lorentzian path integral.
Aiming at going beyond the minisuperspace approximation,
numerical studies of Lorentzian quantum gravity
% using the GTM
have recently been started \cite{Jia:2021xeh,Ito:2022ycc,Jia:2022nda}.
(See Refs.~\cite{Ambjorn:2012jv,Loll:2019rdj} and references therein
for earlier works.)
The importance of using the Lorentzian metric
has also been realized
in nonperturbative formulation of superstring
theory \cite{Kim:2011cr,Nishimura:2019qal}
based on the IKKT matrix model \cite{Ishibashi:1996xs}.
Recent Monte Carlo studies suggest
the emergence of expanding
space-time \cite{Nishimura:2022alt,Anagnostopoulos:2022dak}
unlike in the Euclidean version of the model \cite{Anagnostopoulos:2020xai}.
We consider
that the insights gained in this work
%will be
will 
be important
%have a big impact
% in these directions.
in pursuing these directions further.

Unlike solving the Schr\"odinger equation, the path integral formalism
can be readily extended to many-body systems and field theories
once we overcome the sign problem, for instance, by the GTM
as we have demonstrated.
%We hope our work opens up a new possiblity in that direction.
It should also be emphasized that,
unlike the other promising 
methods \cite{Berges:2005yt,Berges:2006xc,Takeda:2019idb,Takeda:2021mnc},
the GTM has a peculiar advantage
that it is based on the Picard-Lefschetz theory,
which enables us to understand nonperturbative effects
in terms of nontrivial saddle points and the associated thimbles
that appear in evaluating the oscillating integral.
This feature of the GTM was made full use of in our work
in the context of
quantum tunneling,
%the real-time dynamics,
where the connection to semi-classical descriptions was
%obtained.
%particularly important.
of particular importance.
From this point of view, we hope that the GTM is
%particularly useful
also useful
in elucidating various fundamental problems in quantum theory
such as the measurement problem
and 
the quantum-to-classical transition
%its solution
based on the decoherence theory \cite{Schlosshauer:2019ewh}, 
%for instance.
which requires the environment to be included in the simulation.

%% We also think that GTM is also promising
%% work clearly demonstrates the power of the GTM in investigating a system
%% which suffers from severe sign problem.
%% Among various methods that proved promising for that purpose,
%% we consider this method particularly useful
%% in elucidating fundamental problems in quantum mechanics
%% due to the smooth connection to the classical dynamics.

\subsection*{Acknowledgements}

We would like to thank Yuhma Asano and Masafumi Fukuma
%Neill C.\ Warrington
%M.~Honda, Y.~Ito and S.M.~Nishigaki
for valuable discussions.
The computations were carried out on
%the computational resource 
%of KEKCC and the PC cluster in KEK Theory Center.
the PC clusters in KEK Computing Research Center
and KEK Theory Center. K.S. is supported by the Grant-in-Aid for JSPS Research Fellow, No.
20J00079.
A.\ Y.\ is supported by a Grant-in-Aid for Transformative Research Areas
"The Natural Laws of Extreme Universe---A New Paradigm for Spacetime and
Matter from Quantum Information" (KAKENHI Grant No. JP21H05191) from
JSPS of Japan.

\appendix

\section{The calculation method used in this work}
\label{sec:appendix}

In this Appendix, we discuss how we obtained the Monte Carlo results
presented in section \ref{section:complex_traj_from_montecarlo}.
First we briefly review the basic idea of the GTM
to solve the sign problem.
Then we review the idea of integrating the flow time
to solve the multi-modality problem,
and discuss how to apply the HMC algorithm, which enables efficient simulation.
Finally we explain the problem of the anti-holomorphic gradient flow
that occurs in large systems, and discuss how to solve 
%this problem
it by optimizing the gradient flow.

\subsection{the basic idea of the GTM}
%generalized thimble method (GTM)}
\label{section:GTM}

%Monte Carlo method for multidimensional integrals with complex weights is known to be susceptible to the sign problem. Several methods had been proposed to overcome this, including the complex Langevin method \cite{Parisi:1983mgm,Aarts:2009uq,Aarts:2011ax,Scherzer:2019lrh,Scherzer:2018hid,Nagata:2015uga,Nagata:2016vkn,Seiler:2012wz,Ito:2016efb}, the tensor network approach \cite{Levin:2006jai,PhysRevB.86.045139,Adachi:2019paf,Kadoh:2019kqk}, and the Lefschetz thimble method \cite{Picard1897,Lefschetz1924,Witten:2010cx, Witten:2010zr,Fukuma:2020fez, Fujisawa:2021hxh}, among many others. Since it has recently been demonstrated that the thimble method can be used to successfully compute the real-time evolution of the wave function, this approach is thus promising in the study of real-time path integral for quantum tunneling.

In this section we give a brief review of the GTM,
%the Monte Carlo algorithm that is used in our calculation.
which is a promising method for solving the sign problem
based on 
%According to
the Picard-Lefschetz theory.
Here we consider a general model defined by the partition function
and the observable
\begin{equation}
  Z=\int  d^Nx \, e^{-S(x)} \ , \quad
  \langle \mathcal{O} \rangle
  =\frac{1}{Z} \int  d^Nx \, \mathcal{O}(x) \, e^{-S(x)} \ , 
  %\rightarrow\int_{\mathcal{M}}d^Nz\;e^{-S(z)}.
    \label{eq:thible_integral_appendix}
\end{equation}
where $x = (x_1 , x_1 , \cdots , x_N) \in \bbR^N$
and $d^N x = \prod_{n=1}^{N} dx_{n}$.
The action $S(x)$ is a complex-valued holomorphic function of $x$,
which makes \eqref{eq:thible_integral_appendix}
a highly oscillating multi-dimensional integral
and hence causes the sign problem when the number $N$ of variables
becomes large.
This general partition
\eqref{eq:thible_integral_appendix}
includes
%% the real-time path integral
%% \eqref{eq:time_evolution_path_integral-discrete}
%% with the action \eqref{eq:eff-action}
%% as a special case\footnote{This also applies to the more general case
%% %\eqref{eq:time_evolution_path_integral-discrete-Phi}
%%   %with the action
%%   \eqref{eq:eff-action-Phi} with the post-selected wave function $\Phi(x)$
%%   by adding the end point $x_N$ as a new variable.}.
%% %% which also applies to the more general case
the real-time path integral
\eqref{eq:time_evolution_path_integral-discrete-Phi}
with the action \eqref{eq:eff-action-Phi}.
%
%with the post-selected wave function $\Phi(x)$.
%
%can be written in the form \eqref{eq:thible_integral_appendix}.
%The log term in \eqref{eq:eff-action} is 
%\label{footnote:log-action}

Let us first recall that the Picard-Lefschetz theory
makes the oscillating integral well-defined
by deforming the integration
contour
%\eqref{eq:thible_integral_appendix}
using the anti-holomorphic gradient flow equation
\begin{align}
    \frac{d z_i(\sigma)}{d\sigma}
=\overline{\frac{\partial S(z(\sigma))}{\partial z_i}}
\label{anti-hol-grad-flow-general}
\end{align}
%\eqref{anti-hol-grad-flow}
%as 
with the initial condition $z(0)=x\in\mathbb{R}^N$,
where $\sigma$ plays the role of the deformation parameter.
See Eq.~\eqref{anti-hol-grad-flow} and below in section \ref{sec:PL-theory}.
This flow equation defines a one-to-one map
from $x=z(0) \in \bbR^N$ to $z=z(\tau)\in \mathcal{M}_\tau \in \bbC^N$.
Due to Cauchy's theorem, 
the partition function and the observable
\eqref{eq:thible_integral_appendix}
can be rewritten as
\begin{equation}
  Z
  %=\int_{\mathbb{R}^N}d^Nx\;e^{-S(x)}  \rightarrow
  = \int_{\mathcal{M}_\tau}d^Nz \, e^{-S(z)} \ , \quad
  \langle \mathcal{O} \rangle
  =\frac{1}{Z}\int_{\mathcal{M}_\tau}  d^Nz \, \mathcal{O}(z) \, e^{-S(x)} \ . 
    \label{eq:thible_integral_appendix2}
\end{equation}
%% In the $\tau \rightarrow \infty$ limit,
%% ${\rm Im} \, S(z)$ becomes constant on each Lefschetz thimble
%% due to the property \eqref{prop-anti-hol-flow}.
%% Therefore the sign problem is solved except for
%% %the so-called residual sign problem,
%% %which occurs due to
%% the fact that
%% %  the following two things.
%% %  First
%% the integration measure $d^Nz$ is complex
%% in general\footnote{The sign problem due to
%%   the complex integration measure $d^Nz$ is called
%%   the residual sign problem. The severeness of this problem
%%   depends on the model and its parameters. Note also that
%%   it becomes milder by improving the flow equation as we describe in
%%   section \ref{sec:optimized-flow}.}.
%%
%, which makes the integration over each thimble still oscillating.
%%
%% Second the imaginary part ${\rm Im} \, S(z^\star)$ at each
%%   saddle point is different in general, which c. Therefore, if there are
%%   many thimbles that contribute to the integral, 

%% Let us then apply the idea of Monte Carlo methods
%% to \eqref{eq:thible_integral_appendix2}
%% and generate configurations on the deformed contour
%% $\mathcal{M}_\tau$
%% with the probability distribution $\propto e^{-{\rm Re} S(z)}$.
%% Using these configurations,
%% we can calculate
%% the expectation value $\langle \mathcal{O}\rangle$
%% by taking the ensemble average of $\mathcal{O}(z)$
%% with the reweighting factor 
%% $e^{ -i{\rm Im} \, S(z) + i\theta}$,
%% where $\theta$ is the phase of the complex integration measure
%% $d^Nz = |d^Nz| \, e^{i\theta}$.

In the $\tau \rightarrow \infty$ limit,
${\rm Im} \, S(z)$ becomes constant on each Lefschetz thimble
due to the property \eqref{prop-anti-hol-flow}
so that the sign problem is solved except for
the one\footnote{The sign problem due to
  the complex integration measure $d^Nz$ is called
  the residual sign problem. The severeness of this problem
  depends on the model and its parameters \cite{Fujii:2013sra}.}
%%  Whether it can still
%%  be handled by reweighting for large $N$ remains unclear.}
 coming from the measure $d^Nz$.
% In the GTM,
 In the GTM \cite{Alexandru:2015sua},
 the flow time $\tau \rightarrow \infty$ limit is not taken.
 This has a significant advantage over the
 earlier proposals \cite{Fujii:2013sra,Cristoforetti:2012su,Witten:2010cx} 
 with $\tau =\infty$,
 which require 
 prior knowledge of the relevant saddle points.
 %% which has a big advantage over the original proposal with $\tau =\infty$
 %% that prior knowledge of the relevant saddle points is not
 %% required \cite{Alexandru:2015sua}.
%%
%%and hence the integration contour $\mathcal{M}_\tau$ is not decomposed
%%into Lefschetz thimbles.
%%
%This has the virtue of reducing the multi-modality problem
%(or the ergodicity problem) due to 
The sign problem can still be ameliorated by choosing $\tau \sim \log N$,
%large enough
which makes 
the reweighting method work.
%the $\tau \rightarrow \infty$ limit is not taken,
However, the large flow time $\tau$
%there is actually
causes
%the simulation suffers from 
the multi-modality problem
(or the ergodicity problem)
% due to the fact that
since the transitions among different regions of $\mathcal{M}_\tau$
that flow into different thimbles in the $\tau \rightarrow \infty$ limit
are highly suppressed during the simulation.

%\subsection{application of the Hybrid Monte Carlo algorithm}
\subsection{integrating the flow time}
\label{section:integrating-flow-time}

In order to solve both the sign problem and the multi-modality problem,
%at the same time,
it was proposed \cite{Fukuma:2020fez}
to integrate the flow time $\tau$ as
\begin{equation}
  Z_W
  %=\int_{\mathbb{R}^N}d^Nx\;e^{-S(x)}  \rightarrow
  = \int_{\tau_{\rm min}}^{\tau_{\rm max}} d\tau  \, e^{- W(\tau)}
  \int_{\mathcal{M}_\tau}d^Nz \, e^{-S(z)}
  %\ ,
    \label{eq:thible_integral_appendix3}
\end{equation}
%where
with some weight $W(\tau)$, which is chosen to 
make
%%%%%generate the configurations
%%%%$(z_0 , \cdots , z_{N-1}; \tau)$ in such a way that the $\tau$-distribution
%$(z , \tau)$ in such a way that 
the $\tau$-distribution
roughly uniform in the region $ [\tau_{\rm min} , \tau_{\rm max}]$.
The use of this idea is important in our work since we have to
be able to sample all the saddle points and the associated thimbles
that contribute to the path integral.
The validity of our simulation in this regard
is confirmed by reproducing the correct results for the weak value,
which is an ensemble average of the sampled 
trajectories. See the Top panels in Figs.~\ref{fig:ctrl_WM},
\ref{fig:k2_WM}, \ref{fig:fnwf_WM} and \ref{fig:qrtc_WM}.
%For an efficient sampling, it is important to use

%% Once we generate the configurations $(z , \tau)$,
%% we can calculate the expectation value $\langle \mathcal{O} \rangle$
%% by taking the ensemble average $\mathcal{O}(z)$
%% with the reweighting factor using the configurations $(z , \tau)$
%% obtained for an appropriate region of $\tau$.

For an efficient sampling in \eqref{eq:thible_integral_appendix3},
%ween we apply the idea of
we
%it is important to
use the Hybrid Monte Carlo algorithm \cite{Duane:1987de},
which updates the configuration by solving a fictitious classical
Hamilton dynamics treating ${\rm Re} S(z)$ as the potential.
When we apply this idea to \eqref{eq:thible_integral_appendix3},
there are actually two options.
%this fictitious classical Hamilton dynamics.

One option is to
define a fictitious classical Hamilton dynamics for
%define it for
$(z,\tau)\in \mathcal{R}$ with $z \in \mathcal{M}_\tau$,
%$z=z(\tau)$ is the configuration after the flow
where $\mathcal{R}$ is the ``worldvolume'' obtained
by the foliation of $\mathcal{M}_\tau$ with
$\tau \in [\tau_{\rm min} , \tau_{\rm max}]$.
While this option has an important
advantage (See footnote \ref{footnote:modulus}.),
%The disadvantage of this option is that 
one has to treat
a system constrained on the worldvolume,
which makes the algorithm quite complicated.
Another problem is that the worldvolume is pinched if
there is a saddle point on the original integration contour,
which causes the ergodicity problem.
%% This is
%% an issue
%% in the application to the real-time path integral 
%% since we have real saddle points in some cases,
%% which requires us to shift the integration contour in an appropriate way.

Here we adopt the other option, which is to rewrite
\eqref{eq:thible_integral_appendix3} as
\begin{equation}
  Z_W
  = \int_{\tau_{\rm min}}^{\tau_{\rm max}} d\tau  \, e^{- W(\tau)}
   \int d^Nx \,
   \det J(x,\tau)e^{-S(z(x,\tau))} \ ,
   \label{Z-x-parameter-space}
\end{equation}
where 
%$x \mapsto z$.
$z(x,\tau)$ represents the configuration obtained after the flow
starting from $x \in \bbR^N$
and
%$J_{ij}(x , \tau)\equiv \partial z_i(x , \tau) / \partial x_j$
\begin{align}
  J_{ij}(x , \tau) =  \frac{\partial z_i(x , \tau) }{ \partial x_j}
\label{eq:def-Jacobi-matrix}
\end{align}
is the Jacobi matrix associated with the change of variables.
Then one can define a fictitious classical Hamilton dynamics for
$(x,\tau)\in \bbR^N \times [\tau_{\rm min} , \tau_{\rm max}]$.
%% The other is to define it for
%% $(x,\tau)\in \bbR^N \times [\tau_{\rm min} , \tau_{\rm max}]$,
%% where $x=z(0)$ is the configuration before the flow.
Here
%In this second option, 
one only has to deal with an unconstrained
system, which makes the algorithm simple.
The disadvantage, however, is that the Jacobian $\det J(x,\tau)$
that appears in \eqref{Z-x-parameter-space}
has to be taken into
account by reweighting, which causes the
overlap problem\footnote{Note that
  this problem does not occur in the first option
  since the modulus $|\det J(x,\tau)|$ is included in the 
integration measure $|d^Nz|$
in \eqref{eq:thible_integral_appendix3}
although the phase factor $e^{i\theta}= d^N z / |d^N z|$
%\det J(x,\tau)/|\det J(x,\tau)|$
should be taken into account by reweighting.\label{footnote:modulus}}
%% the phase factor of Jacobian $\det J(x,\tau)$,
%% which corresponds to the phase factor $e^{i\theta}$
%% of the measure $d^Nz$,
%% should be taken into account by reweighting, but not
%% the modulus $|\det J(x,\tau)|$.}
when the modulus $|\det J(x,\tau)|$
%of the Jacobian
fluctuates considerably during the simulation.
In that case, only a small number of
configurations with large $|\det J(x,\tau)|$ dominate the
ensemble average and hence the statistics cannot be increased efficiently.
It turns out that this problem does not occur in the simulations performed
in this work if we optimize the flow equation as we describe in
section \ref{sec:optimized-flow}.
In all the simulations, we have chosen $\tau_{\rm min}=0.2$,
which is small enough to solve the multi-modality problem,
and $\tau_{\rm max}=4$,
which is large enough to obtain typical trajectories close to
the relevant saddle points.
Note also that the sign problem is solved already at $\tau \sim 2$.
%%However, the effect of the modulus $|\det J(x,\tau)|$ is already
%%included in the fictitous classical dynamics

Once we generate the configurations $(x , \tau)$,
we can calculate the expectation value $\langle \mathcal{O} \rangle$
by taking the ensemble average of $\mathcal{O}(z(x,\tau))$
with the reweighting factor 
%$\det J(x,\tau) \, e^{i \, {\rm Im}S(z(x,\tau))}$
\begin{align}
R(x,\tau) = \det J(x,\tau) \, e^{- i \, {\rm Im}S(z(x,\tau))}
%e^{ -i{\rm Im} \, S(z) + i\theta} \ ,
\label{eq:reweithting-factor}
\end{align}
using the configurations
%$(z(x,\tau) , \tau)$
$( x , \tau)$
obtained for an appropriate range of $\tau$ \cite{Fukuma:2021aoo}.

In either option of the HMC algorithm, the most time-consuming
part is the calculation of the Jacobian $\det J(x,\tau)$,
which 
requires O($N^3$) computational cost
%is needed only 
in the reweighting procedure.
In order to calculate the Jacobi matrix $J(x,\tau)$,
%The Jacobian $J(x;\tau)\equiv \partial z/\partial x$ can be computed
one has to solve the flow equation
\begin{equation}
  \frac{\partial}{\partial \sigma}J_{ij}(\sigma)
%  =\overline{\frac{\partial^2 S(x,\sigma)}{\partial z_i\partial z_k}\;
  =\overline{H_{ik}(z(\sigma))\, J_{kj}(\sigma)}
    \label{eq:Jacobian_floweq}
\end{equation}
with the initial condition $J(0)=\textbf{1}_N$,
where we have defined the Hessian
\begin{align}
  H_{ij}(z)
  &= \frac{\partial^2 S(z)}{\partial z_i\partial z_j} \ .
  \label{eq:def-Hessian}
\end{align}

\subsection{backpropagating Hybrid Monte Carlo algorithm}
\label{sec:backpropagation}

In this section we review the backpropagating HMC algorithm
%proposed in Ref.~\cite{Fujisawa:2021hxh}.
\cite{Fujisawa:2021hxh},
which is crucial in simulating the system \eqref{Z-x-parameter-space}.
%% the fictitious classical Hamilton dynamics for
%% $(x,\tau)\in \bbR^N \times [\tau_{\rm min} , \tau_{\rm max}]$.
%which is proposed in Ref.~\cite{Fujisawa:2021hxh}.
Here we discuss the case of fixed flow time $\tau$
%in \eqref{Z-x-parameter-space}
for simplicity
and comment on the case of integrating $\tau$ at the end of this section.

The first step of the HMC algorithm \cite{Duane:1987de}
is to introduce new variables
$p_i$ ($i=1, \cdots , N$) with the partition function
\begin{equation}
  Z_\text{HMC}=\int
  % _{\mathbb{R}^N\times \mathbb{R}^N}
  d^Nx \, d^Np \, e^{-\frac{p^2}{2}-\text{Re}S(z(x,\tau))} \ ,
  \label{eq:Z-HMC}
\end{equation}
where the Gaussian integral of $p$ simply yields a constant factor.
%which is equivalent to the original model.
In order to update the configuration $(x,p)$,
we first generate $p$ with the probability distribution $\propto \exp(-p^2/2)$
and solve the fictitious Hamilton equation with the Hamiltonian
\begin{equation}
    H(x,p)= \frac{1}{2} \sum_{i=1}^N p_i^2 + \text{Re}S(z(x,\tau)) \ ,
\end{equation}
which reads
\begin{align}
  \label{Ham-eq-x}
  \frac{dx_i(s)}{ds} &=    p_i(s) \ , \\
  \frac{dp_i(s)}{ds} &=  F_i(s)
  % \ ,
\label{Ham-eq-p}
\end{align}
with the force $F_i(s)$ defined by
\begin{align}
  F_i(s) &= - \left. \frac{\del {\rm Re}S(z(x,\tau))}{\del x_i} \right|_{x=x(s)}
  \ .
\label{def-F}
\end{align}
%%
%% \begin{alignat}{3}
%%   H(x,p)= \frac{1}{2} \sum_{i=1}^N p_i^2 + \text{Re}S(z(x,\tau)) \ ,
%% \end{alignat}
%%
We solve the Hamilton equation
\eqref{Ham-eq-x} and \eqref{Ham-eq-p}
for a fixed time $s_{\rm f}$ to obtain
a new configuration $x(s_{\rm f})$ and $p(s_{\rm f})$.

In actual calculation, we discretize the Hamilton equation
using the standard leap-frog discretization,
which respects the reversibility and the preservation
of the phase space volume \cite{Duane:1987de}.
Let us 
%We introduce the $\Delta s$ and 
divide the total time $s_{\rm f}$ into $N_{\rm s}$ segments
as $s_{\rm f} = N_{\rm s} \Delta s$.
Then we define the discretized Hamilton equation as
\begin{align}
    x_i(s_{n+1/2})&=x_i(s_n)+\frac{\Delta s}{2} \, p_i(s_{n}) \ , \\
    p_i(s_{n+1})&=p_i(s_n)+\Delta s \, F_i(s_{n+1/2}) \ , \\
    x_i(s_{n+1})&=x_i(s_{n+1/2})+\frac{\Delta s}{2} \, p_i(s_{n+1})
    %\ ,
\end{align}
for $n=0,1,...,N_s-1$, 
where $s_\nu = \nu \Delta s$ with $\nu$ being an integer or a half integer.
Since the Hamiltonian conservation is violated by the discretization,
we have to treat the new configuration
given by $x(s_{\rm f})$ and $p(s_{\rm f})$
as a trial configuration, which is 
subject to the Metropolis accept/reject procedure
with the acceptance probability
$\min (1 , e^{- \delta H})$, where
\begin{align}
\delta H & = H(x(s_{\rm f}),p(s_{\rm f})) - H(x(0),p(0))  \ ,
\label{dela-H}
\end{align}
which guarantees the detail balance exactly.
The parameters $s_{\rm f}$ and $N_{\rm s}$ in the HMC algorithm can be
optimized in a standard way by minimizing the computational cost
required for generating a statistically independent configuration.

Note that the force \eqref{def-F} can be rewritten as
\begin{equation}
  F_i(s)
  %=-\left.\frac{\partial}{\partial x_i}\text{Re}S(z(x,\tau))\right|_{x=x(s)}
  =
  % 2 \text{Re}\left\{
  f_j(s)
%  \frac{\partial S(z(x(s),\tau))}{\partial z_j}
  J_{ji}(x(s),\tau) 
  %   \right\} \ ,
  +   \overline{f_j(s)   J_{ji}(x(s),\tau) } \ ,
  \label{eq:naive-formula-force}
\end{equation}
where we define the ``force'' 
\begin{align}
f_i(s) &= - \left. \frac{\del {\rm Re} S(z)}{\del z_i} \right|_{z=z(x(s),\tau)}  
\label{def-f-deformed}
\end{align}
at $z(x(s),\tau)$ on the deformed contour $\mathcal{M}_\tau$.
%, which is obtained after the flow starting from $x(s)$.
If we use 
%this formula 
eq.~\eqref{eq:naive-formula-force}
to calculate the force,
we need to calculate the Jacobi matrix $J_{ji}(x(s),\tau)$
at each step of solving the Hamilton equation.
It was found recently \cite{Fujisawa:2021hxh} that this can be avoided
by using backpropagation as we explain below.

Let us first rewrite the flow equations
for the configuration \eqref{anti-hol-grad-flow-general}
and the Jacobi matrix \eqref{eq:Jacobian_floweq}
in the discretized form as\footnote{The original version
of this argument was given in Appendix A of Ref.~\cite{Fujisawa:2021hxh}
without discretizing the flow equations.}
\begin{align}
  \label{eq:flow-z-conf}
  z_i(\sigma+\epsilon)&=
  z_i(\sigma)
  %+\epsilon\overline{\partial_i S(\sigma)},\\
  +\epsilon \, \overline{\frac{\partial S(z(\sigma))}{\partial z_i}} \ ,
  \\
  J_{ij}(\sigma+\epsilon)&=J_{ij}(\sigma)
  %+\epsilon\overline{\partial_i\partial_j S(\sigma)}\overline{J_{kj}(\sigma)}.
  + \epsilon \, \overline{ H_{ik}(z(\sigma)) J_{kj}(\sigma)} \ .
  \label{eq:flow-jacob-mat}
\end{align}
%Here, $\partial\partial S(\sigma)$ means evaluating the Hessian at $z=z(\sigma)$.
Note that
\eqref{eq:flow-jacob-mat}
%the flow equation for the Jacobi matrix
can be written in a matrix notation as
\begin{align}
  \mat{J(\sigma+\epsilon)\\\overline{J(\sigma+\epsilon)}}
  &=
  %\underset{\displaystyle\mathcal{U}(\sigma)}
%  {
%    \underbrace{
\mathcal{U}(\sigma)
  \mat{J(\sigma)      \\      \overline{J(\sigma)}
  } \ ,
  \label{eq:J-flow-matrix}
  \\
%  \quad
  \mathcal{U}(\sigma)
%  = \mat{\textbf{1}_N & \epsilon\overline{\partial\partial S(\sigma)}
  &= \mat{\textbf{1}_N & \epsilon \overline{H(z(\sigma))}
      \\
%      \epsilon\partial\partial S(\sigma)&\textbf{1}_N }   \ .
      \epsilon  H(z(\sigma))&\textbf{1}_N }   \ .
    \label{eq:def-Umat}
\end{align}
Using this, we can rewrite the force \eqref{eq:naive-formula-force} as
%in terms of $\mathcal{U}(\sigma)$ as
\begin{align}
  F^\top(s)
  &=
  %  \mat{\partial S(\tau)&\overline{\partial S(\tau)}}
  \mat{ f^\top(s) & \bar{f}\, {}^{\top}(s) }
  %\overline{f}^\top(s)}
  \mat{J(\tau)\\\overline{J(\tau)}}\nonumber\\
  &=
%  -\frac{1}{2}
  %  \mat{\partial S(\tau)&\overline{\partial S(\tau)}}
  \mat{ f^\top(s) &  \bar{f}\, {}^{\top}(s) }
%    \overline{f}^\top(s)}
  \mathcal{U}(\tau-\epsilon) \, \mathcal{U}(\tau-2\epsilon)\cdots\mathcal{U}(2\epsilon) \, \mathcal{U}(\epsilon)\mat{\textbf{1}_N\\\textbf{1}_N} \ .
  \label{eq:UUU-expression}
\end{align}
Note that evaluating this quantity by multiplying the matrices
from the right corresponds to
using 
eq.~\eqref{eq:naive-formula-force}
%the formula \eqref{eq:naive-formula-force}
naively to calculate the force.
This requires matrix-matrix multiplications,
which cost O($N^3$) computation time
%arithmetic operations 
or O($N^2$) if the Hessian is sparse
as in local systems.
However, we can actually evaluate \eqref{eq:UUU-expression}
%this quantity
by multiplying the matrices from the left,
which corresponds to backpropagating the force on the deformed contour
to the original contour.
This requires only vector-matrix multiplications
and thus reduces the computational cost by the order of O($N$)
in this procedure.

%% If one were to evaluate this quantity by multiplying the matrices
%% from the right (forward propagation),
%% which is what is traditionally done,
%% we cannot avoid the repeated operation of matrix-matrix multiplications,
%% which is of order O($N^3$) in complexity (or O($N^2$) for local systems).
%% But since $\mat{\partial S(\tau)&\overline{\partial S(\tau)}}$ is a vector,
%% multiplying the matrices from the left (backward propagation)

In our simulation, we actually use the optimized flow equation
explained in the next section,
and hence the flow equations \eqref{eq:flow-z-conf}
and \eqref{eq:flow-jacob-mat} have to be modified accordingly.
However, the idea of backpropagation
remains applicable.
% can still be used.

So far we have explained the idea of the HMC algorithm
for a fixed flow time $\tau$ for simplicity.
In actual simulation, however, we also integrate the flow time $\tau$
as in \eqref{Z-x-parameter-space} to avoid the multi-modality problem.
%% Note also that the flow time $\tau$ should also be integrated
%% as in \eqref{Z-x-parameter-space}.
Accordingly, we have to treat $\tau$ as
a dynamical variable in the HMC algorithm
together with its conjugate momentum $p_\tau$.
The partition function \eqref{eq:Z-HMC} should then be replaced by
\begin{align}
  \tilde{Z}_{\rm HMC}
  \label{Z-HMC-W-mtau}
%=\int_{M_\tau} dz \, e^{-S(z)}
  &=\int d\tau \, dp_\tau \, dx  \,  dp  \, e^{-H}  \ , \\
  H &= \frac{1}{2 m(\tau)} (p_i)^2 + \frac{1}{2} (p_\tau)^2
  + {\rm Re} S(z(x,\tau)) + W(\tau) \ ,
\label{def_Ham-W-mtau}
\end{align}
where we introduce the $\tau$-dependent mass function $m(\tau)$.
%and refer the readers to
See
section 6.1 of
Ref.~\cite{Fujisawa:2021hxh} for the details.
%in the case of integrating $\tau$.
Here we just mention that $m(\tau)$ should be chosen to be
proportional to 
the typical value of $|\det J(x,\tau)|^{2/N}$
for various $x$
%on the deformed contour
with fixed $\tau$
so that the simulation can realize
a random walk on the deformed manifold with 
almost uniform discretization.
Also the weight function $W(\tau)$ in
\eqref{Z-x-parameter-space}
should be chosen so that the distribution of $\tau$
obtained by simulations becomes as uniform as possible within the
region $\tau_{\rm min} \le \tau \le \tau_{\rm max}$.
%$\tau \in [\tau_{\rm min}, \tau_{\rm max}]$.
The functions $m(\tau)$ and $W(\tau)$ used in each simulation
are parametrized as
\begin{align}
m(\tau) &= m_0 + m_1  \tau  \ , \\ 
W(\tau) &= \sum_{k=1}^6 w_k \tau^k \ .
\label{eq:W-function}
\end{align}
We use $m_0=-0.209$ and $m_1=1.5851$ for all cases,
while $w_k$ are chosen as in Table \ref{tab:W}.
\begin{table}
    \centering
    \begin{tabular}{|c||c|c|c|c|c|c|}
        \hline
Figure & $w_1$ & $w_2$ & $w_3$ & $w_4$ & $w_5$ & $w_6$ 
        \\
        \hline
        Fig.3 & -26.4297 & 25.8099 & -13.814 & 4.0988
         & -0.6254 & 0.0382 \\
        Fig.4 & -31.949 & 35.409 & -21.3925 & 7.0716
         & -1.1934 & 0.0805 \\
        Fig.5 & -33.9717 & 36.9599 & -21.7341 & 7.0366
         & -1.1676 & 0.0775 \\
        Fig.6 & -26.344 & 27.6191 & -15.0622 & 4.5638
         & -0.7146 & 0.0452 \\
        Fig.7(TL) & -26.4127 & 25.5518 & -13.7184 & 4.0885
         & -0.6254 & 0.0382 \\
        Fig.7(TR) & -26.4127 & 25.5518 & -13.7184 & 4.0885
         & -0.6254 & 0.0382 \\
        Fig.7(BL) & -27.748 & 30.1805 & -18.2515 & 6.04706
         & -1.0202 & 0.06868 \\
        Fig.7(BR) & -64.38 & 68.644 & -40.113 & 12.9891
         & -2.1781 & 0.14779 \\
        \hline
    \end{tabular}
    \caption{The parameters \eqref{eq:W-function}
      in the function $W(\tau)$ chosen for the case shown in each figure.
      TL, TR, BL and BR in the left-most column imply Top-Left, Top-Right,
      Bottom-Left and Bottom-Right, respectively.}
    \label{tab:W}
\end{table}

As for the parameters in the HMC algorithm,
we always use $s_{\rm f}=1$, whereas
$N_{\rm s}$ is chosen to be $10$
except for the cases in Fig.~\ref{fig:classical_limit},
%corresponding to $\hbar=0.5$,
where we use $N_{\rm s}=30$ to keep the acceptance rate high enough.

%\subsection{Thimble method with optimized flow equation}
\subsection{optimizing the flow equation}
\label{sec:optimized-flow}

In this section, we discuss a problem\footnote{See
  Ref.~\cite{Feldbrugge:2022idb} for discussions on the anti-holomorphic
  gradient flow and its modification from a different point of view.} 
that occurs when we use
the original flow equation 
\eqref{anti-hol-grad-flow-general}
%\eqref{eq:original_flow_eq}
%that exists appears
for a system with many variables such as 
\eqref{eq:time_evolution_path_integral-discrete-Phi}
%\eqref{eq:thible_integral_appendix}
with $N=20$ studied in this paper.
%with many dynamical degrees of freedom.
We solve this problem by optimizing the flow equation, which
actually has large freedom of choice
if we are just to satisfy the property \eqref{prop-anti-hol-flow}.
%and its resolution.
Here we explain the basic idea and
defer a detailed discussion to the forth-coming paper \cite{precondition2022}.
%A full-detail analysis is given in Ref. \cite{precondition2022}.

The problem with the original flow
\eqref{anti-hol-grad-flow-general}
can be readily seen by considering how its solution $z(x,\sigma)$ changes
when the initial value $z(x,0) = x \in \bbR^N$ changes infinitesimally.
Note that the displacement 
%Let us define
$\zeta_i(\sigma)\equiv z_i(x+\delta x,\sigma)-z_i(x,\sigma)$
for an infinitesimal $\delta x$
%% to be a displacement coming from difference of the
%% initial configurations $\delta$.
%%With a sufficiently small $\delta$, this vector
can be obtained as
\begin{align}
\zeta_i(\sigma) = J_{ij}(\sigma) \, \delta x_j \ ,
\end{align}
where $J_{ij}(\sigma)$ is the Jacobi matrix at the flow time $\sigma$,
which satisfies the flow equation \eqref{eq:Jacobian_floweq}.
Thus we find that the displacement satisfies the flow equation
\begin{equation}
  \frac{d\zeta_i(\sigma)}{d\sigma}
  %  =\overline{\frac{\partial^2S(z(x,\sigma)}
    =\overline{H_{ij}(z(\sigma)) \, \zeta_j(\sigma)}
\label{eq:zeta-time-evolve}
\end{equation}
with the boundary condition $\zeta_i(0)=\delta x_i$,
where $H_{ij}(z)$ is the Hessian defined by \eqref{eq:def-Hessian}.

Let us consider the singular value decomposition (SVD)
of the Hessian $H_{ij}(z(\sigma))$ given as\footnote{This is known
  as the Takagi decomposition,
  which is the SVD for a complex symmetric matrix.}
\begin{align}
  H(z(\sigma)) = U^\top (\sigma) \, \Lambda(\sigma) \, U(\sigma) \ ,
\label{eq:H-SVD}
\end{align}
where $U(\sigma)$ is a unitary matrix and
$\Lambda={\rm diag}(\lambda_1 , \cdots , \lambda_N)$
is a diagonal matrix with $\lambda_1 \ge \cdots \ge  \lambda_N \ge 0$.
Plugging this in \eqref{eq:Jacobian_floweq},
we obtain
\begin{align}
  \frac{d J (\sigma)}{d\sigma}
  %  =\overline{\frac{\partial^2S(z(x,\sigma)}
  = U(\sigma)^\dag \, \Lambda(\sigma) \,
  \overline{U(\sigma) \, J(\sigma)} \ ,
\label{eq:Jacobian-time-evolve2}
\end{align}
and similarly for the displacement
%() \eqref{eq:zeta-time-evolve}, we obtain
\begin{align}
  \frac{d\zeta(\sigma)}{d\sigma}
  %  =\overline{\frac{\partial^2S(z(x,\sigma)}
  = U(\sigma)^\dag \, \Lambda(\sigma) \,
  \overline{U(\sigma) \, \zeta(\sigma)} \ .
\label{eq:zeta-time-evolve2}
\end{align}
%% When the condition number $\eta (H) \equiv \lambda_1 / \lambda_N$ is large,
%%   the magnitude of the right-hand side \eqref{eq:zeta-time-evolve2} can
%%   be very different for different $\delta x_i$.
  Roughly speaking, the magnitude of the
  displacement $\zeta(\sigma)$  grows exponentially
  with $\sigma$, and
  the growth rate
  %  the rate of the exponential growth
  is given by
  a weighted average of the singular values with a weight
  depending on $\delta x$.
If the singular values have a hierarchy $\lambda_1 \gg \lambda_{N}$,
some modes grow much faster than the others.
This causes a serious technical problem in solving the flow equation 
  since it may easily diverge during the procedure.
%when one updates the configuration during the simulation.

  In order to solve this problem, we pay attention to the freedom in
  defining the flow equation. As we discussed in section
  \ref{sec:PL-theory}, the important property of
  the flow equation \eqref{anti-hol-grad-flow-general}
  is \eqref{prop-anti-hol-flow}.
  Let us therefore consider a generalized flow equation
\begin{align}
    \frac{d z_i(\sigma)}{d\sigma}
    =\mathcal{A}_{ij}(z(\sigma),\overline{z(\sigma)})
      \overline{\frac{\partial S(z(\sigma))}{\partial z_j}} \ ,
\label{anti-hol-grad-flow-general-opt}
\end{align}
%is a function of $z$ and $\bar{z}$.
which generalizes
%Then 
the equation \eqref{prop-anti-hol-flow}
% becomes
as
\begin{equation}
    \frac{d S(z(\sigma))}{d\sigma}
=\sum_i \frac{\partial S(z(\sigma))}{\partial z_i}
    \frac{d z_i(\sigma)}{d \sigma} 
    =\sum_{ij} \frac{\partial S(z(\sigma))}{\partial z_i}
    \mathcal{A}_{ij}(z(\sigma),\overline{z(\sigma)})
    %(z(\sigma),\bar{z}(\sigma))
\overline{\frac{\partial S(z(\sigma))}{\partial z_j}} \ .
%\overline{\frac{\partial S_{\rm eff}(z(\sigma))}{\partial z}}
%\geq 0 \ ,
\label{prop-anti-hol-flow-general}
\end{equation}
For this to be positive semi-definite,
the kernel $\mathcal{A}_{ij}(z,\bar{z})$ has only to be Hermitian positive,
and it does not have to be holomorphic.\footnote{While
this generalization does not change the saddle points,
it changes the shape of the thimbles associated with them.
Note, however, that the integral over each thimble
% associated with each saddle point remains 
remains unaltered due to Cauchy's theorem.}
%Note that it can be a non-holomorphic function of $z$.
%is  does $\mathcal{A}$ is not necessarily holomorphic.

Accordingly,
%The 
the flow of the Jacobi matrix
becomes
\begin{equation}
  \frac{\partial}{\partial \sigma}J_{ij}(\sigma)
%  =\overline{\frac{\partial^2 S(x,\sigma)}{\partial z_i\partial z_k}\;
  =\mathcal{A}_{ik} \overline{H_{kl}(z(\sigma))\, J_{lj}(\sigma)}
  + \left(
  \frac{\partial \mathcal{A}_{il}}{\partial z_k} J_{kj}(\sigma)
  + \frac{\partial \mathcal{A}_{il}}{\partial \bar{z}_k} \overline{J_{kj}(\sigma)} \right)
    \overline{ \frac{\partial S(z(\sigma))}{\partial z_l}} 
 \ .
      \label{eq:Jacobian_floweq-preconditioned}
\end{equation}
%where the last term appears since $\mathcal{A}$ is not necessarily holomorphic.
Note that the discretized version of \eqref{eq:Jacobian_floweq-preconditioned}
can still be written in the form \eqref{eq:J-flow-matrix},
which means that the backpropagation \cite{Fujisawa:2021hxh} can be used
even with the generalized flow equation.

From \eqref{eq:Jacobian_floweq-preconditioned},
we obtain the flow of the displacement as
\begin{equation}
  \frac{\partial}{\partial \sigma} \zeta_i(\sigma)
%  =\overline{\frac{\partial^2 S(x,\sigma)}{\partial z_i\partial z_k}\;
  =\mathcal{A}_{ik} \overline{H_{kl}(z(\sigma))\, \zeta_l(\sigma)}
  + \left(
  \frac{\partial \mathcal{A}_{il}}{\partial z_k} \zeta_k(\sigma)
  + \frac{\partial \mathcal{A}_{il}}{\partial \bar{z}_k} \overline{\zeta_k(\sigma)} \right)
    \overline{ \frac{\partial S(z(\sigma))}{\partial z_l}} 
 \ .
      \label{eq:zeta_floweq-preconditioned}
\end{equation}
Let us here assume that the first term is
dominant\footnote{This assumption is valid when $z(\sigma)$ is
  close to a saddle point, for instance. Otherwise, it should be 
simply regarded as a working hypothesis.}
in \eqref{eq:zeta_floweq-preconditioned}.
Then plugging \eqref{eq:H-SVD}
%\eqref{eq:zeta-time-evolve} 
in \eqref{eq:zeta_floweq-preconditioned},
we obtain
\begin{align}
  \frac{d\zeta(\sigma)}{d\sigma}
  %  =\overline{\frac{\partial^2S(z(x,\sigma)}
  \sim \mathcal{A} \, U(\sigma)^\dag \, \Lambda(\sigma) \,
  \overline{U(\sigma) \, \zeta(\sigma)} \ .
\label{eq:zeta-time-evolve2-opt}
\end{align}
Therefore, by choosing 
%which suggests
\begin{align}
  \mathcal{A} = U(\sigma)^\dag \, \Lambda^{-1}(\sigma)\, U(\sigma) \ ,
  \label{eq:A-optimal}
  \end{align}
we 
%can make \eqref{eq:zeta-time-evolve2} as
obtain
\begin{align}
  \frac{d\zeta(\sigma)}{d\sigma}
  %  =\overline{\frac{\partial^2S(z(x,\sigma)}
  \sim U(\sigma)^\dag \, \overline{U(\sigma) \, \zeta(\sigma)} \ ,
\label{eq:zeta-time-evolve3}
\end{align}
in which 
%Thus we can eliminate 
the problematic hierarchy of
singular values $\lambda_i$ in \eqref{eq:zeta-time-evolve2}
%\eqref{eq:zeta-time-evolve2}.
is completely eliminated.
From this point of view, \eqref{eq:A-optimal} seems to be the optimal choice
for the ``preconditioner'' $\mathcal{A}$ 
in the generalized
%anti-holomorphic 
flow equation \eqref{anti-hol-grad-flow-general-opt}.
Note also that, under a similar assumption, the flow of the Jacobi matrix
changes from \eqref{eq:Jacobian-time-evolve2} to
\begin{equation}
  \frac{\partial}{\partial \sigma}J(\sigma)
%  =\overline{\frac{\partial^2 S(x,\sigma)}{\partial z_i\partial z_k}\;
  \sim
   U(\sigma)^\dag \, \overline{U(\sigma) \, J(\sigma)} \ .
      \label{eq:Jacobian_floweq-preconditioned-2}
\end{equation}
Thus, in our simulation,
the use of the optimal flow equation
solves
%is expected also to reduce 
the overlap problem that actually occurs otherwise
due to the large fluctuation of $|\det J|$. (See the discussions
below \eqref{eq:def-Jacobi-matrix}.)
%algorithm.
%in our method based on \eqref{Z-x-parameter-space}.

In order to implement this idea,
% in the simulation, 
let us first note 
that \eqref{eq:A-optimal} can be written as
\begin{align}
  \mathcal{A}(z(\sigma),\overline{z(\sigma)}) 
  %  A(z,\bar{z})
  &=
  \Big\{ H^\dag(z(\sigma)) H(z(\sigma)) \Big\}^{-1/2}
  = \Big\{ \overline{H(z(\sigma))} H(z(\sigma)) \Big\}^{-1/2} \ .
  %% \mathcal{A} &=   A(z(\sigma),\overline{z(\sigma)}) \ , \\
  %% A(z,\bar{z}) &=
  %% \Big\{ H^\dag(z) H(z) \Big\}^{-1/2}
  %% = \Big\{ \overline{H(z)} H(z) \Big\}^{-1/2} \ .
%  = (\bar{H} \bar{H}^\dag)^{-1/2} \ .
  \label{eq:A-optimal-2}
  \end{align}
Here we use the rational
approximation
%% \footnote{These techniques are
%%   widely used with the HMC algorithm, which is known
%%   as the rational HMC
%%   algorithm \cite{Kennedy:1998cu,Clark:2006wq}.
%% %% ,Clark:2006wp}.
%%   Its application includes
%%   QCD with dynamical strange quarks \cite{Clark:2004cp} and
%%   supersymmetric theories such as the BFSS and IKKT matrix 
%% models (See Refs.~\cite{Kim:2011cr,Anagnostopoulos:2007fw}, for example.).} 
\begin{align}
x^{-1/2} 
&\approx
a_0  + \sum_{q=1}^Q \frac{a_q}{ x + b_q } \ ,
\label{eq:rational_approximation}
\end{align}
which can be made accurate for a wide range of $x$
%to the extent one needs
with the real positive parameters $a_q$ and $b_q$
generated by the Remez algorithm.
Thus we obtain
\begin{align}
    \mathcal{A}(z,\bar z)
    &\approx a_0 \, \textbf{1}_N
    + \sum_{q=1}^Q a_q \, \Big\{ \overline{H(z)} H(z)+b_q\, \textbf{1}_N \Big\}^{-1} \ .
    \label{eq:rational_approximation-2}
\end{align}
With this expression, the derivative of $\mathcal{A}$
in \eqref{eq:Jacobian_floweq-preconditioned} can be calculated
straightforwardly as
\begin{align}
  \frac{\partial\mathcal{A}}{\partial z_k}
  &=-\sum_{q=1}^Qa_q (\bar HH+b_q\textbf{1}_N)^{-1}\bar H 
  \frac{\partial H}{\partial z_k}
  (\bar HH+b_q\textbf{1}_N)^{-1} \ ,  \\
  \frac{\partial \mathcal{A}}{\partial \bar{z}_k}
  &=-\sum_{q=1}^Qa_q(\bar HH+b_q\textbf{1}_N)^{-1}
  \overline{ \frac{\partial H}{\partial z_k} }
  H(\bar HH+b_q\textbf{1}_N)^{-1}  \ . 
\end{align}
%Note that the flow equation \eqref{eq:Jacobian_floweq-preconditioned}
The matrix inverse $(\bar HH+b_q\textbf{1}_N)^{-1}$ does not have to
be calculated explicitly since it only appears in the algorithm 
as a matrix that acts on a particular vector, which allows us to
use an iterative method for solving a linear equation
such as the conjugate gradient (CG) method.
The factor of $Q$ in the computational cost
can be avoided by the use of a multi-mass CG solver \cite{Jegerlehner:1996pm}.
These techniques are well known 
in the so-called Rational HMC
  algorithm \cite{Kennedy:1998cu,Clark:2006wq},
%% ,Clark:2006wp}.
which is widely used in
%  Its application includes
  QCD with dynamical strange quarks \cite{Clark:2004cp} and
  supersymmetric theories such as the BFSS and IKKT matrix 
models (See Refs.~\cite{Catterall:2007fp,Anagnostopoulos:2007fw,Kim:2011cr}, 
for example.). 
%% These techniques are
%%   widely used with the HMC algorithm, which is known
%%   as the rational HMC
%%   algorithm \cite{Kennedy:1998cu,Clark:2006wq}.
%% %% ,Clark:2006wp}.
%%   Its application includes
%%   QCD with dynamical strange quarks \cite{Clark:2004cp} and
%%   supersymmetric theories such as the BFSS and IKKT matrix 
%% models (See Refs.~\cite{Kim:2011cr,Anagnostopoulos:2007fw}, for example.). 

%% The matrix inverse $(\bar HH+b_q\textbf{1}_N)^{-1}$ does not have to
%% be calculated explicitly since it only appears in the algorithm 
%% as a matrix that acts on a particular vector, which allows us to
%% use an iterative method for solving a linear equation
%% such as the conjugate gradient method.

%% which is much easier to handle on a computer
%% than \eqref{eq:A-optimal}.
%% In particular, one can use the conjugate gradient (CG) method in
%% acting the matrix inverse in \eqref{eq:rational_approximation-2}
%% on a vector. The factor of $Q$ in the computational cost
%% can be avoided by the use of a multi-mass CG solver \cite{Jegerlehner:1996pm}.
%% These techniques are
%%   widely used with the HMC algorithm, which is known
%%   as the rational HMC
%%   algorithm \cite{Kennedy:1998cu,Clark:2004cp,Clark:2006wq,Clark:2006wp}.
%%   Its application includes
%%   QCD with dynamical strange quarks and
%%   supersymmetric theories such as the BFSS and IKKT matrix models

%\bibliographystyle{unsrt}
\bibliographystyle{JHEP}
\bibliography{ref}

\end{document}